\documentclass[submit]{aa}  

\usepackage{graphicx}
\usepackage{txfonts}
\usepackage{natbib}
\begin{document}

\title{Fast fitting of spectral lines with Gaussian and hyperfine structure models}

\author{Mika Juvela, Devika Tharakkal}

\institute{                                  
Department of Physics, P.O.Box 64,
FI-00014, University of Helsinki, Finland,
{\em mika.juvela@helsinki.fi}
}

\authorrunning{M. Juvela et al.}  

\date{Received September 15, 1996; accepted March 16, 1997}

\abstract
{The fitting of spectral lines is a common step in the analysis of line
observations and simulations. However, the observational noise, the presence
of multiple velocity components, and potentially large data sets make it a
non-trivial task.}
{We present a new computer program Spectrum Iterative Fitter (SPIF) for the
fitting of spectra with Gaussians or with hyperfine line profiles. The aim is
to show the computational efficiency of the program and to use it to examine
the general accuracy of approximating spectra with simple models.}
{We describe the implementation of the program. To characterise its
performance, we examined spectra with isolated Gaussian components or a
hyperfine structure, also using synthetic observations from numerical
simulations of interstellar clouds. We examined the search for the globally
optimal fit and the accuracy to which single-velocity-component and
multi-component fits recover true values for parameters such as line
areas, velocity dispersion, and optical depth. }
{The program is shown to be fast, with fits of single Gaussian components
reaching on graphics processing units speeds approaching one million spectra
per second. This also makes it feasible to use Monte Carlo simulations or
Markov chain Monte Carlo calculations for the error estimation. However, in
the case of hyperfine structure lines, degeneracies affect the parameter
estimation and can complicate the derivation of the error estimates. }
{The use of many random initial values makes the fits more robust, both
for locating the global $\chi^2$ minimum and for the selection of the optimal
number of velocity components.}

\maketitle

\section{Introduction}

Observations of spectral lines are typically analysed by fitting the data with
simplified models, and this is also true for the molecular and atomic emission
lines observed at radio wavelengths, where the Gaussian approximation is the
most common one. The simplest model can thus consist of a single Gaussian that
has three free parameters (central velocity, velocity dispersion, and the peak
value). The Gaussian model can in principle be justified by the assumption of
a Maxwellian velocity distribution of the gas, if the velocity dispersion is
dominated by thermal motions. More generally, the central limit theorem
suggests that with enough random motions within the beam (e.g. turbulence),
the shape of the observed spectrum approaches a Gaussian. However, the
observed spectra are rarely precisely Gaussian, and they can show asymmetries
(e.g. due to rotation or infall), contain both a narrow and a wide component
(e.g. due to outflows or shocks), or contain completely distinct velocity
components due to the line-of-sight alignment of separate gas clouds (e.g. large
clouds or observations at low Galactic latitudes). The complexity is often
addressed by fitting multiple Gaussians, although the physical interpretation
of an individual component can be less clear. However, the sum of the fitted
components approximates the observed line profile, and such multi-component
fits can be used simply as a tool to estimate quantities such as the peak
intensities, line areas, and velocity dispersions.  Also the column densities
are typically estimated based on the fits rather than by direct integration of
the relevant quantities over individual channels, for which the
signal-to-noise ratio (S/N) is much worse.

Fits to hyperfine spectra follow the same pattern, combining several Gaussians
with fixed velocity offsets and usually with fixed relative strengths. The 
hyperfine fits still use Gaussians to represent the components of the
underlying optical depth, but the fits are also directly concerned with the
physical parameters of excitation temperature $T_{\rm ex}$ and the total
optical depth $\tau$. There is a qualitative difference to the simpler
Gaussian fits above, where the fit is an empirical description of the line
intensity profile and, in principle, it could even include a sum of positive and
negative Gaussians that individually have no physical interpretation.  Similar
freedom does not exist in hyperfine fits where a single unphysical velocity
component can also lead to the end result being unphysical.

Although the fitting of both Gaussian and hyperfine structure models is in
principle technically straightforward, some challenges still exist. These are
partly computational, caused by the large line surveys and especially
numerical simulations that can produce millions of spectra. When visual
inspection of all fits is no longer feasible, one needs methods that are both
fast and robust. One factor in the robustness is the selection of the initial
parameter values for the optimisation of the fitted model. In non-linear least
squares problems, and especially when spectra contain multiple components, it
is not guaranteed the optimisation would converge to a globally optimal
solution. Some optimisation algorithms, such as simulated annealing or genetic
algorithms,
are more likely to find the global $\chi^2$ minimum, but they also tend to be
computationally more expensive. The determination of the optimal number of
components becomes another important consideration, which may need to take
several factors into account, beyond just the formal goodness of the fit. 

Pyspeckit \citep{Ginsburg2011, Pyspeckit22}, SCOUSEPY \citep{Scousepy16,
Scousepy19}, GaussPy \citep{Lindner2015}, GaussPy+ \citep{Gausspy19}
BTS \citep{Clarke2018}, and MWYDIN \citep{Rigby2024} 
are some of the software packages used in the analysis of radio spectral
lines.
GaussPy is the Python implementation of the autonomous Gaussian decomposition
(AGD) algorithm discussed in \cite{Lindner2015} and has been used extensively 
in the 21-SPONGE survey \citep{Murray2017, Murray2018}. GaussPy decomposes any
spectra that can be modelled using Gaussian functions and utilises a machine
learning algorithm to find the appropriate smoothing parameters for the data.
Once trained, GaussPy can decompose around 10 000 spectra with Ncpus (number
of central processing units, CPUs) in approximately 3/Ncpu hours \citep[see
Appendix~D]{Lindner2015}. Recently, a fully automated package GaussPy+ was
designed based on the GaussPy algorithm and a comparative study of these two
programs is given in \citet{Gausspy19}. GaussPy and GaussPy+ do not support
more complicated spectral decomposition such as a hyperfine structure. 
Pyspeckit is another CPU parallelised Python-based spectral fitting tool
\citep{Ginsburg2011, Pyspeckit22}, which includes a range of spectral model
functions (Gaussian, Lorentzian, and Voight) and ready-to-use model types ($\rm
NH_3$, $\rm N_2H+$, $\rm HCN$, $\rm ^{13}CO$, and $\rm C^{18}O$). It has been
used recently for example in the ChaMP survey to analyse the hyperfine structure
of HCN \citep{2017MNRAS.465.2559S}. Various other spectral decomposition
pipelines, including astroclover \citep{Zeidler2021} and pyspecnest
\citep{Sokolov2020}, also utilise the fitting tools in Pyspeckit. SCOUSEPY (a
Python interface of the program SCOUSE written in IDL \citep{Scousepy16})
uses the cube fitting module of Pyspeckit. \citet{Scousepy19} provides a brief
description of the spectral statistics using SCOUSEPY where they analysed
around 300,000 pixels and, after smoothing, modelled around 130,000 spectra with a 96.4\%
success \citep{Scousepy19}.
BTS \citep{Clarke2018} selects the number of fitted components by analysing
the first three derivatives of the intensity (versus velocity), once the
spectrum has been smoothed to reduce the effect of observational noise. The
number of components is selected automatically based on the reduced $\chi^2$
values of the alternative fits. Finally, the program MWYDYN \citep{Rigby2024}
is geared towards the automatic fitting of hyperfine spectra with up to three
velocity components. The program uses the standard assumptions of a common
excitation temperature and full width at half maximum (FWHM) for all hyperfine
components, which are also assumed to have Gaussian profiles. The number of
components is selected based on Bayesian information criterion (BIC). The code
also checks the neighbouring pixels for better fits and uses those iteratively
as initial values for alternative fits.

In this paper we describe a new computer program, Spectral iterative fitter
(SPIF). The computational challenges of large sets of spectra are met by using
parallelisation and graphics processing units (GPUs) to speed up the fitting.
This allows one to address the problem of initial values in a general way, by
simply repeating the fits a number of times with different initial values. It
also has become possible to estimate the uncertainty of the fitted parameters
with Monte Carlo methods, even for spectral cubes consisting of millions of
spectra. In addition to describing the implementation and the basic
characteristics of SPIF, we analyse synthetic molecular line spectra,
including some more realistic examples from numerical cloud simulations. We
use the results to characterise the precision to which the basic Gaussian and
hyperfine spectrum models are likely to describe the complexity of real
observations.

The contents of the paper are the following. In
Section~\ref{sect:implementation} we describe the implementation of the SPIF
program and the spectral models that are being fitted. The calculations behind
the synthetic observations are described in Sect.~\ref{sect:MHD}. The results
are presented in Sect.~\ref{sect:results}. We examine there the computational
performance of SPIF in the case of Gaussian fits
(Sect.~\ref{sect:gaussian_fits} and hyperfine fits (Sect.~\ref{sect:hfs}).
Section~\ref{sect:precision} examines how well the fitted spectral models are
able to describe the spectra from the cloud simulation, and the question of
error estimates is studied separately in Sect.~\ref{sect:error_estimates}. We
discuss the results in Sect.~\ref{sect:discussion} before presenting the
conclusions in Sect.~\ref{sect:conclusions}.

\section{Implementation} \label{sect:implementation}

The SPIF program can be used to fit spectra with one or more Gaussian
components or a hyperfine structure. For $N$ Gaussians, the fitted model is
$T_{\rm A}$,
\begin{equation}
{\hat T}_{{\rm A},i} =  \sum_{k=1}^{N}  
T_k \exp [-4 \ln 2  \left(  \frac{v_i-v_{k}}{\Delta v_{k}} \right)^2  ],
\label{eq1}
\end{equation}
where the free parameters are $T_{k}$ for the peak value, $\Delta v_k$ for the
full width at half maximum (FWHM) of the Gaussian, and $v_k$ for the central
velocity. The index $k$ refers to the fitted component, and $i$ represents an
individual velocity channel. The solution is found by minimising the $\chi^2$
value
\begin{equation}
\chi^2  =  \sum_{i=1}^{M}  
\left(  
           \frac{T_{\rm A,i}-{\hat T}_{\rm A,i}}{\delta T_{A}}    
\right)^2,
\end{equation}
where the sum is over $M$ velocity channels. The error estimate $\delta
T_{\rm}$ is assumed to be the same for all channels in a spectrum. We use
later also the quantity $\chi^2_{\rm N}$ (reduced $\chi^2$) that is obtained
by dividing $\chi^2$ with the degree of freedom.

In the case of hyperfine spectra, the optical depth is first calculated as the
sum over individual hyperfine components, 
\begin{equation}
\tau_i   =  \sum_{k=1}^{}  \tau \times I_k 
           \exp [-4 \ln 2  \left(  \frac{v_i-v-v_{k}}{\Delta v_{k}} \right)^2  ],
\end{equation}
where $I_{k}$ and $v_k$ are the relative opacities and velocity offsets of
the hyperfine components and $v$ is the radial velocity. The model for the
antenna temperatures $T_{\rm A}$ is
\begin{equation}
{\hat T}_{\rm A,i}^{\rm pred} =  [ J(T_{\rm ex})-J(T_{\rm bg})] \times (1-e^{-\tau_i}).
\label{eq2}
\end{equation}
Here $J$ is 
\begin{equation}
J(T) = \frac{h \nu/k}{\exp(h \nu/(k T))-1},
\end{equation}
$T_{\rm bg}$ is the assumed temperature of the source background, and $\nu$ is
the frequency of the transition. The above relation assumes a single velocity
component, resulting in four free parameters: the excitation temperature
$T_{\rm ex}$, the velocity (usually defined as the radial velocity of one of
the hyperfine components), the FWHM line width, and the optical depth $\tau$. 

SPIF consists of a Python host program and a set of kernels, which are
compiled programs implemented using
OpenCL\footnote{https://www.khronos.org/opencl/} libraries. At the core of the
SPIF program are the optimiser kernels. These allow the calculations to be
performed either on the host computer (using just the central processing unit
CPU) or alternatively on a graphics processing unit (GPU). The latter provides
access to massive parallelisation that is well suited for line fitting, when
the inputs consist of a large number (preferably thousands) of spectra that
can be fitted independently of each other. We have implemented three
optimisers, a naive componentwise gradient descent routine, the Nelder-Mead
Simplex algorithm, and a conjugate gradient optimiser. Of these, the gradient
descent was intended mainly for initial testing but has turned out to be
relatively fast for simple problems. The Simplex and he gradient descent
method (in spite of its name) use only the $\chi^2$ and no gradient
information. This also enables the use arbitrary penalty functions\footnote{
A penalty function can in principle be any expression containing the
optimised parameter. Some examples are included in Appendix~\ref{app:manual}.}
or to easily extend SPIF with new spectral models. In the case of the
conjugate gradient algorithm, the derivatives relative to the parameters of
the Gaussian and the hyperfine models are calculated analytically.
Calculations are performed by default in single precision. This is typically
on GPUs much faster than the use of double precision. On the other hand, the
limited precision could cause problems if derivatives (including those
associated with the penalty functions or priors) were calculated numerically.

The user provides SPIF with an initialisation file that lists the FITS files
with the spectral cubes for the observations and their error estimates.  For
hyperfine fits, an additional input file is used to specify the frequency of
the transition and the relative velocities $v_i$ and intensities $I_i$ of the
hyperfine components. The initialisation file includes further the description
of the fitted model, any potential penalty functions that are used, and the
instructions for the initialisation of the optimised parameters. The initial
parameters can be constants, calculated based on the spectra (e.g. using the
intensity and velocity of the maximum emission or the average velocity of the
emission) or they can be read from an external file as separate parameters for
each spectrum. 
The options for initial values are discussed further in
Appendix~\ref{app:manual} and also in Sect.~\ref{sect:test_gaussian}.
At the run time, the model specification is added to the OpenCL
kernel code, which is compiled on the fly for the actual calculations. This
results in a flexible but computationally efficient system. For example, the
penalty functions (or priors) can be any arbitrary c-language expressions that
depend on the optimised parameters, global constants, or values read from
auxiliary files (individual values for each spectrum). In this paper we use
only penalty functions of the form
\begin{equation}
\Delta \chi2  = \left( \frac{y_0-y}{\delta y} \right)^2, \, \,{\rm if} \, \, y<y_0,
\label{eq:penalty}
\end{equation}
where $y$ refers to a fitted parameter and $y_0$ is a constant threshold value
(in this case a lower limit), and $\delta y$ is a constant that specifies the
steepness of the penalty. SPIF also allows the simultaneous fitting of spectra
read from two input files, such as different transitions observed towards the
same sky position. This then allows for further constraints, such as using the
same radial velocity for both spectra in a pair.

\begin{figure}
\sidecaption
\includegraphics[width=8.8cm]{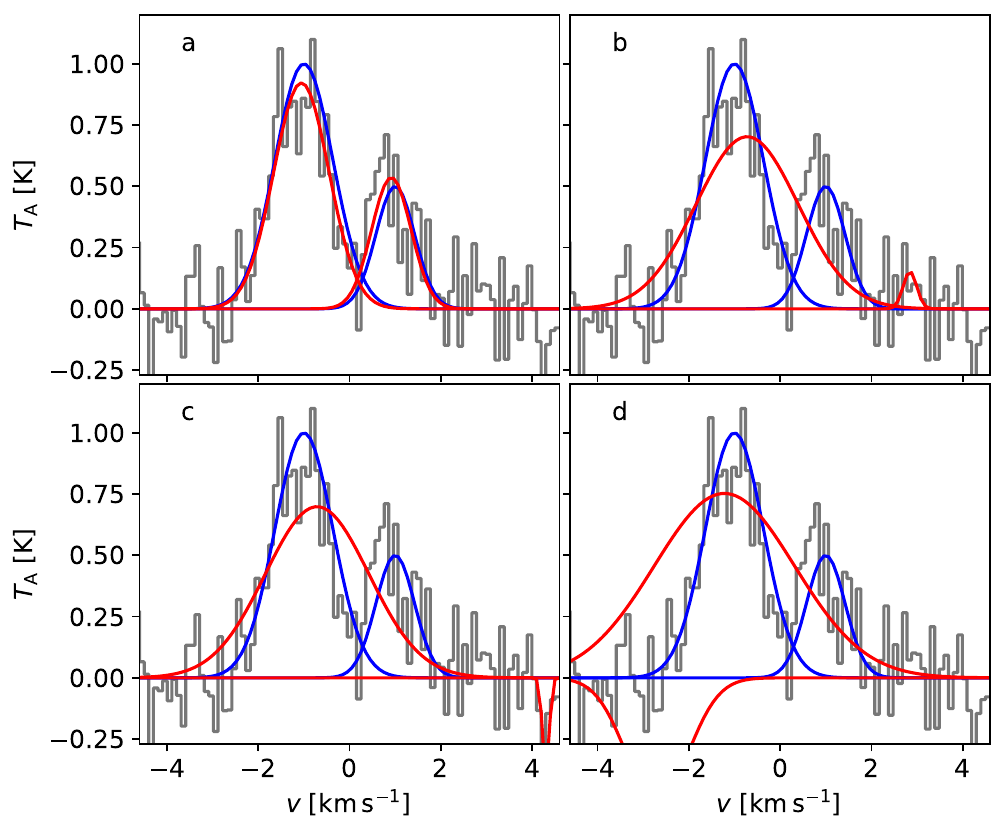}
\caption{
Effect of initial values of the model parameters. The frames show a synthetic
spectrum (black histogram) that correspond to two Gaussian peaks (solid blue
lines) and white noise. The frames shows four potential outcomes, where
$\chi^2$ minimisation with different initial values has led to different
outcomes. The solid red lines show the fitted two components.  The fits
correspond to local $\chi^2$ minima, when no other constraints are used.
}
\label{fig:fails}
\end{figure}

SPIF includes different types of iterations, where each spectrum is fitted
multiple times. Once the number of fitted components is selected, a fit
can be repeated $N_{\rm iter}$ times with perturbed initial parameter values,
and the program will return the results from the fit that resulted in the
lowest $\chi^2$ value. A non-linear least-squares problem (or more generally
non-linear optimisation, allowing for arbitrary penalty functions) can have
several $\chi^2$ minima. Thus, a fit can converge to a local minimum that
depends on the initial values chosen for the free parameters.
Figure~\ref{fig:fails} illustrates the potential problem in the case of just
two Gaussian components and noise. The fit may pick some noise peak (frames
b-d) or, without further constraints, lead to potentially unphysical solutions
with negative components (frames c-d).  With large enough $N_{\rm iter}$ and
large enough variation in the initial values (also defined in the
initialisation file), SPIF should be able to find the solution corresponding
to the global $\chi^2$ minimum, but other constraints may also be needed.

The uncertainty of the fitted model parameters can be estimated with Monte
Carlo simulation. In this iteration the observed $T_{\rm A}$ values are
perturbed according to their error estimates. This is repeated $N_{\rm MC}$
times, the distribution of the fitted parameter values providing the
information of the uncertainties and correlations between the model
parameters. The previous iteration types can also be combined. The first
iteration (original spectra fitted with $N_{\rm iter}$ random initial values)
provides the estimate for the model parameters, and these are then used as the
initial values in the fitting of $N_{\rm MC}$ Monte Carlo realisation of the
input spectra. However, the fit can be repeated with different initial values
even for each of the $N_{\rm MC}$ Monte Carlo realisations. This increases the
probability that the fit to each Monte Carlo sample is also the optimal fit
for that noise realisation.

SPIF includes some basic Markov chain Monte Carlo routines, which provide an
alternative way to calculate error estimates. However, these are less
reliable, because the Markov chains may sometimes show poor mixing. This is
partly due to the special challenges with the simulated observations examined
in this paper, where the peak $T_{A}$ values of noiseless spectra can
vary over many orders of magnitude. Therefore, in the following we concentrate
mainly on the use of the Monte Carlo error estimates.

\section{Synthetic spectra} \label{sect:MHD}

We use in the tests synthetic line observations that are based on the
magnetohydrodynamic (MHD) simulations of a star-forming cloud presented in
\citep{Haugbolle2018} and are similar to the data used in
\citet{Juvela2022_ngVLA}. The MHD model covers a volume of (4\,pc)$^3$ with
octree discretisation providing a maximum linear resolution of 100\,au or
$4.88 \times 10^{-4}$\,pc. We use a (1.26\,pc)$^3$ sub-volume that contains
the highest densities. The original MHD cube (with periodic boundary
conditions) was rotated to bring the most massive cloud filament to the centre
of the model, before selecting the final sub-volume. 

The production of synthetic line maps started with the modelling of the dust
temperature distribution. The radiation field consists of the external
interstellar radiation field \citep{Mathis1983} and of stellar sources 
produced as part of the MHD simulation, as described in
\citet{Juvela2022_ngVLA}. The continuum radiative transfer program SOC
\citep{Juvela2019_SOC} was used to solve the dust temperatures for each cell
in the model. In the absence of direct information on the gas kinetic
temperatures $T_{\rm kin}$, the dust temperature was then used as a proxy for
$T_{\rm kin}$. This is well justified at high densities, $n({\rm H_2}) \ga
10^5$\,cm$^{-3}$ \citep{Goldsmith2001,JuvelaYsard2011a}. The modelling of the
spectral lines made further use of the density and velocity fields of the MHD
simulation. The line maps were produced with the non-LTE radiative transfer
program LOC \citep{Juvela2020_LOC}. The peak fractional abundances $x_0$ were
set to $2\times 10^{-6}$ for $^{13}$CO, $2\times 10^{-7}$ for ${\rm C^{18}O}$,
and $1\times 10^{-10}$ for ${\rm N_2 H^{+}}$. 
We used an additional density scaling \citep[cf.][]{Glover2010}, such
that the final density-dependent abundances are
\begin{equation}
x(n({\rm H}_2)) =  x_0 \frac{n({\rm H}_2)^{2.45}}{3.0\times 10^8+n({\rm H}_2)^{2.45}}.
\end{equation}
We calculated the $^{13}$CO(1-0), C$^{18}$O(1-0), and ${\rm N_2 H^{+}}$(1-0)
spectra, which in the case of ${\rm N_2 H^{+}}$ took into account the full
hyperfine structure of the $J=1-0$ transition. The background temperature
is $T_{\rm bg}=2.73$\,K. Further details on the calculations can be found in
\citep{Juvela2022_ngVLA}

The radiative transfer calculations resulted in maps for three orthogonal view
directions, each with 2576$\times$2576 pixels. These corresponds to the
highest spatial resolution of the MHD model. However, because of the
hierarchical discretisation, the true resolution is lower outside the densest
structures. The velocity resolution of the extracted spectra was set to
0.1\,km\,s$^{-1}$ for the $^{13}$CO(1-0) and C$^{18}$O(1-0) and
0.2\,km\,s$^{-1}$ for the ${\rm N_2 H^{+}}$ spectra.  The total bandwidth is
12\,km\,s$^{-1}$) for the CO lines and 44.8\,km\,s$^{-1}$ for ${\rm N_2
H^{+}}$. In the tests we used a series of maps where the number of pixels was
further reduced by a factor of $R$ per dimension, once the data were first
convolved with a Gaussian beam with $FWHM$ equal to $2R$ pixels.

\begin{figure*}
\sidecaption
\includegraphics[width=12cm]{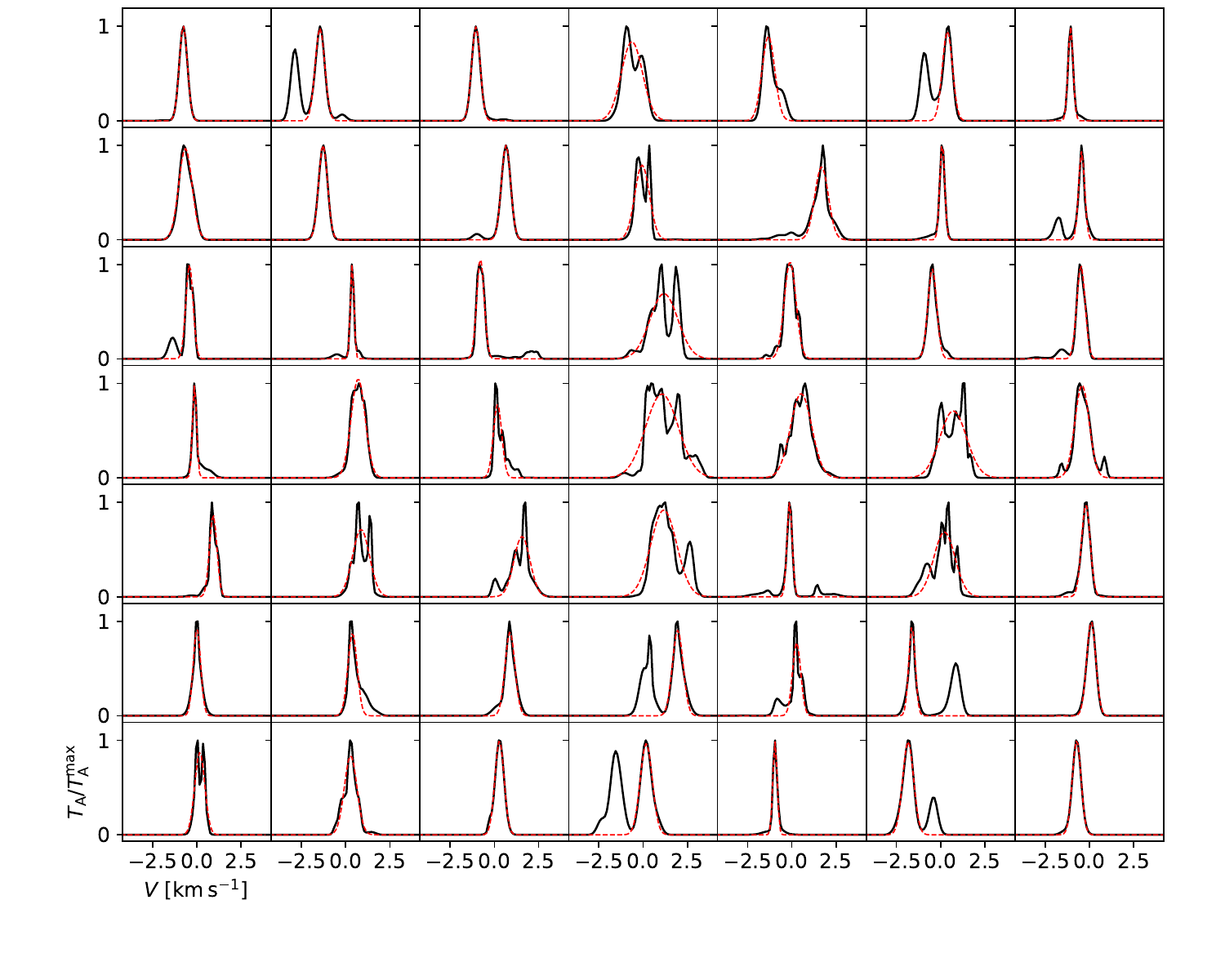}
\caption{
Examples of fits to synthetic spectra from the MHD model (solid black lines) 
with a single Gaussian component (dashed red lines). The spectra are for
C$^{18}$O (direction $x$, $R$=8) with no added observational noise. The
initial values for the intensity and velocity were set based on the peak
values in the spectra. The plotted spectra were normalised to a maximum
value of one. 
}
\label{fig:spectra}
\end{figure*}

\section{Results} \label{sect:results}

In this section we examine fits of Gaussian and hyperfine spectra with the
SPIF program. In addition to computational performance
(Sect.~\ref{sect:performance}), we are interested in how well the Gaussian
fits generally reproduce the spectra (Sect.~\ref{sect:precision}). This is
tested with the help of the data described in Sect.~\ref{sect:MHD}.  Some
characterisation of the synthetic observations can be found in
Appendix~\ref{app:stat}. We finish by looking at the error estimation, mainly
with the Monte Carlo simulations (Sect.~\ref{sect:error_estimates}).

\subsection{Computational performance} \label{sect:performance}

\subsubsection{Gaussian fits} \label{sect:gaussian_fits}

Figure~\ref{fig:spectra} presents examples of the C$^{18}$O spectra from the
MHD cloud model described in Sect.~\ref{sect:MHD},  without added
observational noise. It shows how the spectra often contain several velocity
components, although only up to three distinct peaks among the examples in the
figure (cf. Appendix~\ref{app:stat}). The SPIF fits are started using the
velocity and intensity of the maximum emission in each spectrum and a fixed
$FWHM=$1\,km\,s$^{-1}$ line width. Figure~\ref{fig:spectra} shows examples of
the outcome when complex spectra are fitted with a small number of velocity
components: the solution may converge to a single peak (not necessarily the
strongest one) or the fitted profile can become wide, matching the sum of
multiple peaks.

\begin{figure}
\sidecaption
\includegraphics[width=8.8cm]{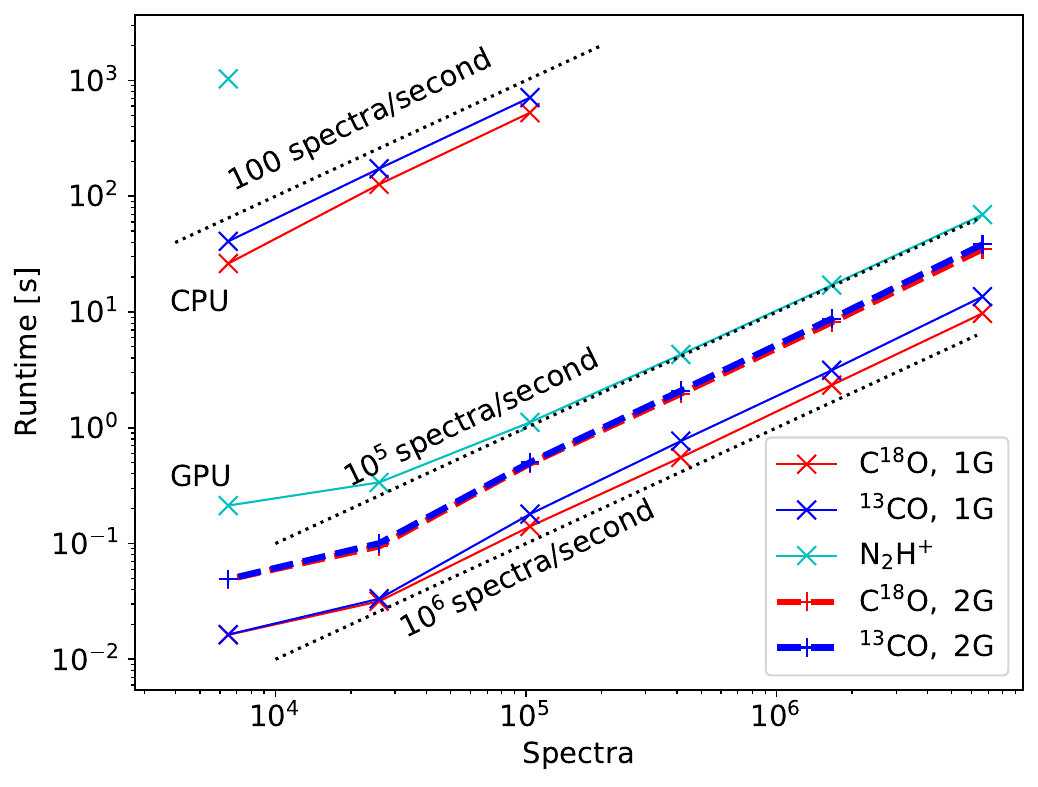}
\caption{
Runtimes of Gaussian and hyperfine fits. Results are shown for Gaussian fits
to C$^{18}$O (red lines) and $^{13}$CO (blue lines) spectra and for hyperfine
fits to $\rm N_2H^{+}$ spectra (cyan lines). The CPU speeds are slightly below
(i.e. faster) than the dotted line for 100 spectra per second. The GPU fits
are in the range of $10^5-10^6$ spectra per second. In the legend, 1G and
2G refer to fits with one or two Gaussian components, respectively. The timings
correspond to the $x$ view direction, but the runtimes are nearly identical
for the other view directions.
}
\label{fig:timings}
\end{figure}

Figure~\ref{fig:timings} shows the runtimes for one- and two-component
Gaussian fits to C$^{18}$O and $^{13}$CO data.
In this case, no additional observational noise was added to the spectra
obtained from the radiative transfer modelling. 
The fits were done in using single precision and the $R$=1-32 maps, where the
total number of spectra ranges from 6480 ($R=32$) to to 6.6 million ($R=1$).
The values are for the actual fit and do not include the cost of reading the
observations from disk and storing of the results. These are, however, a minor
part in the overall runtimes. The timings in Fig.~\ref{fig:timings} are for
the naive gradient descent algorithm. The runtimes of the conjugate gradient
algorithm would be similar, while the Simplex method tends to be a few times
slower. The fits used all the 120 velocity channels, although significant
emission typically covers a smaller fraction of the full bandwidth. 

There is a large difference in the speed of the CPU and GPU runs, the latter
being faster by almost three orders of magnitude. The {\tt leastsq} algorithm
of the Scipy library\footnote{www.scipy.org} would reach a speed of some 40
spectra per second, and the CPU speed in Fig.~\ref{fig:timings} is roughly in
line with that, these SPIF run using four CPU cores. Both the CPU and GPU
runtimes scale linearly with the number of spectra, and the GPU efficiency
drops only for the smallest samples, where the number of spectra falls below
the number of physical computing units on the GPU (a top level consumer-grade
GPU). There was no significant difference between the C$^{18}$O and $^{13}$CO
runtimes.

We next look at the distribution of the $\chi^2$ values as a function of
noise, the number of fitted Gaussian components, and the number of iterations
when fits are repeated with different random initial values. The data
correspond to the $R=1$ maps of C$^{18}$O and the combination of all three
orthogonal view directions. We remove from the sample spectra where the peak
emission remains below 10\,mK, leaving about 4.67 million spectra. In the fits
with two and three velocity components, we apply additional penalties for
negative intensities (Eq.~(\ref{eq:penalty}) with $y=T_{\rm A}$, $y_0=0$\,K,
and $\delta y=0.01$\,K) and small line widths ($y=FWHM$,
$y_0=0.05$\,km\,s$^{-1}$, and $\delta y=0.01$\,km\,s$^{-1}$). These reduce the
probability for unphysical solutions in the case of multi-component fits, such
as spectra decomposed into the sum of arbitrarily large positive and negative
components. It also prevents the appearance of very narrow Gaussian components
that would fall between velocity channels and could correspondingly have
arbitrarily high peak values.
Many of our synthetic spectra have intensities that are smaller than the
selected values of $\delta y$, resulting in only a small effect from the
penalty. A constant $\delta y$ is more appropriate for real observations, when
all spectra have a similar absolute noise level.
The use of penalty functions had no noticeable
effect on the runtimes, which are naturally directly proportional to the
number of iterations.

\begin{figure}
\sidecaption
\includegraphics[width=8.8cm]{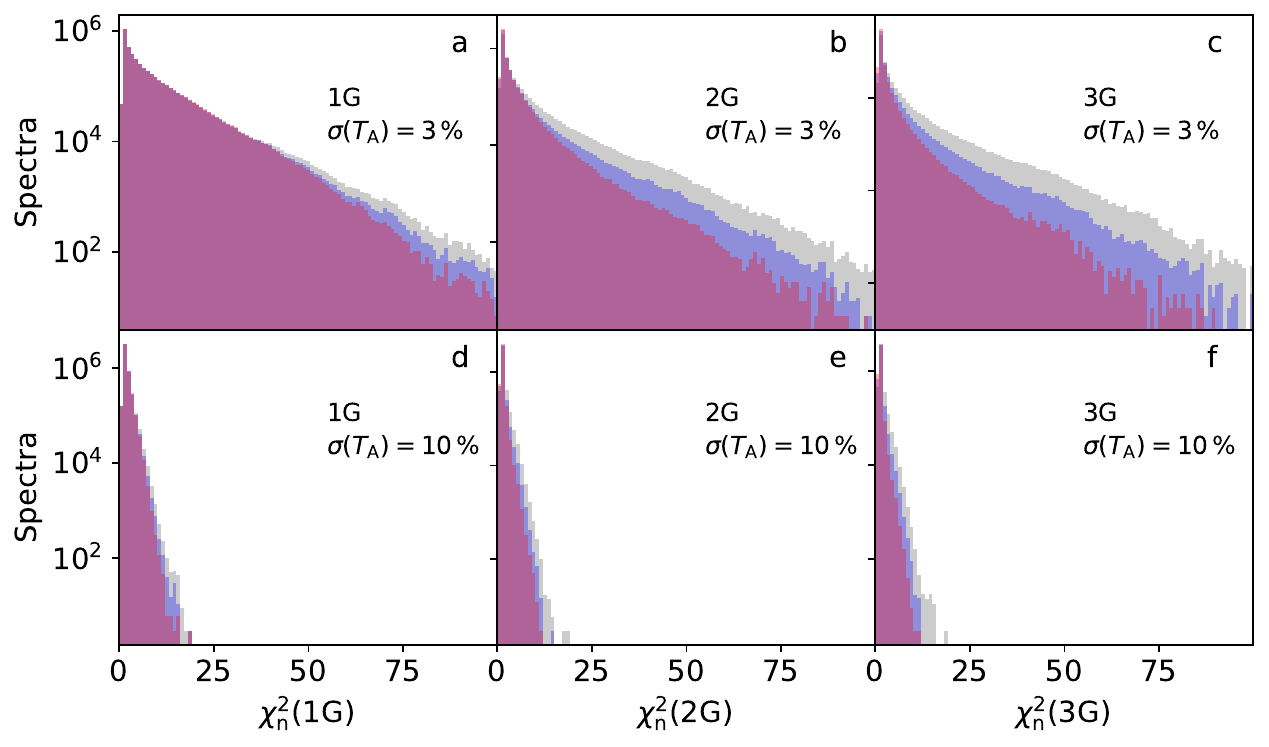}
\caption{
Distributions of $\chi^2_{\rm N}$ values the fits of C$^{18}$O spectra with
Gaussian components. The noise in the upper frames is 3\% of the maximum value
of a spectrum and 10\% in the lower frames. Results are shown for fits with
one, two, or three Gaussian components (frames from left to right), and for
$N_{\rm iter}$=1, 5, and 20 (grey, blue, and red histograms, respectively). 
}
\label{fig:chi2}
\end{figure}

Figure~\ref{fig:chi2} shows the $\chi^2_{\rm N}$ distributions for 1-3
component Gaussian fits with two noise levels and after $N_{\rm iter}$=1, 5,
and 20 iterations. The noise $\sigma(T_{\rm A})$ was set equal to either 3\%
or 10\% of the maximum value of each spectrum, making the S/N the same for all
spectra. The initial values on the first iteration were set according to
velocity and intensity of the spectrum maximum, with $FWHM=1$\,km\,s$^{-1}$.
On subsequent iterations, the initial value were drawn from a normal
distribution with $\sigma$ equal to 30\% around the values on the first
iteration. 
There is a large difference in the normalised $\chi^2$ values between the 3\%
and 10\% noise cases, caused by other emission that remains unexplained by the
fitted model. With increasing number of retries, the recovered best $\chi^2$
values can naturally only decrease. The reduction is noticeable even after
five retries, although the improvement is mostly only at 10\% level. When the
noise is increased to 10\% of the peak value, the $\chi^2$ values are
significantly lower, but the fraction improved fits is not very different from
the $\sigma(T_{\rm A})$=3\% case.

\subsubsection{Hyperfine structure fits} \label{sect:hfs}

\begin{figure*}
\sidecaption
\includegraphics[width=12cm]{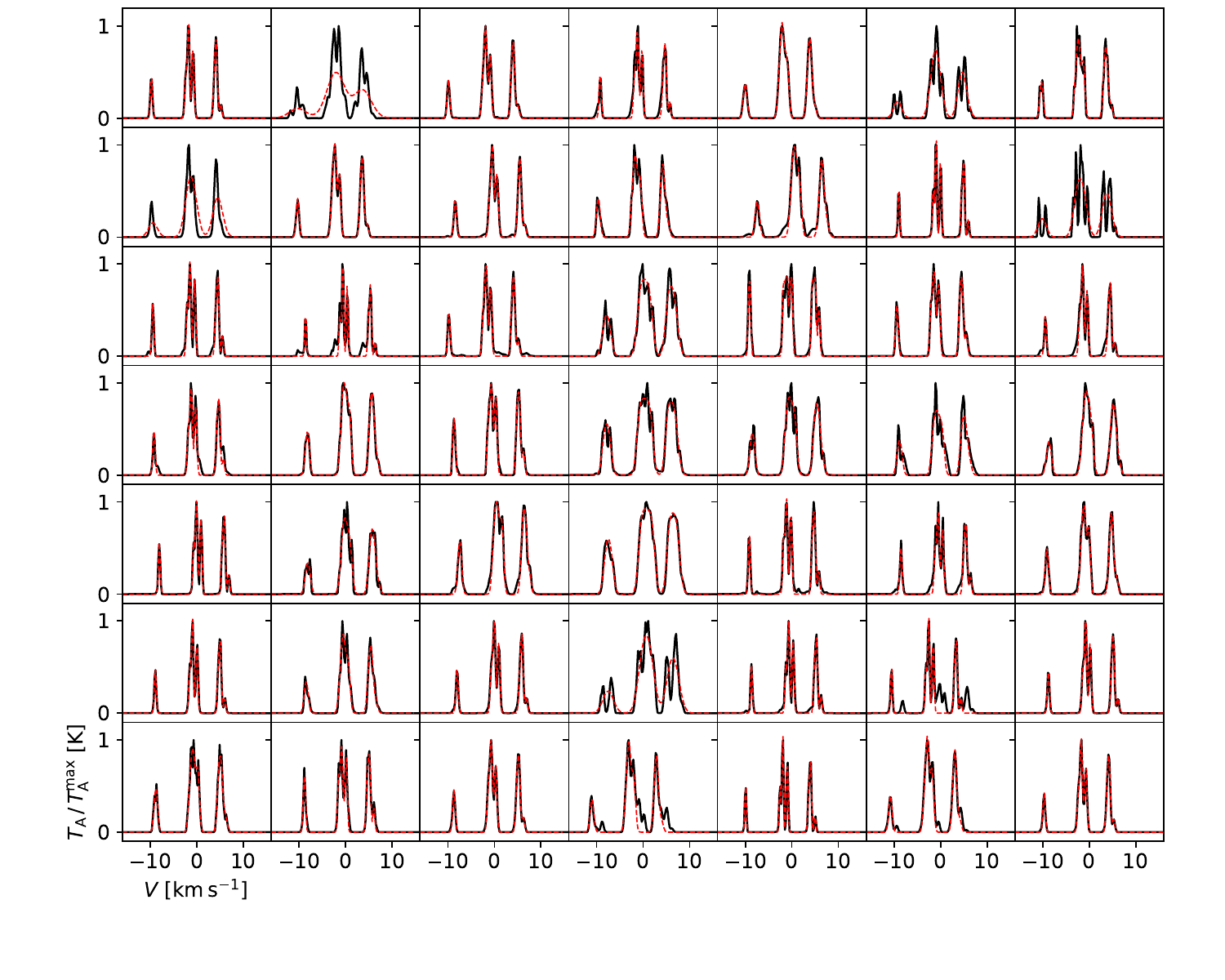}
\caption{
Examples of synthetic ${\rm N_2H^{+}}$ spectra (solid black lines) and the
fitted hyperfine profiles (dashed red lines). The spectra are for direction
$x$, $R$=8, and without added observational noise. The y-axis of each frame was
scaled independently, and the shown $T_{\rm A}$ axis values apply to the
leftmost frames only.
}
\label{fig:n2h+_spectra}
\end{figure*}

The fits to ${\rm N_2H^{+}}$ hyperfine spectra were timed using a single
velocity component. The fits involved four free parameters and included
penalties for small excitation temperatures $T_{\rm ex}\le T_{\rm bg}$ and
negative $\tau$ values. However, in this case the penalty functions had almost
no effect on the results. We examined spectra without added noise and with
$T_{\rm A}$ error estimates equal to 3\% of the peak value of each spectrum.
The initial values for the optimisation were $T_{\rm ex}$=5\,K, $v$ equal to
the mean velocity of the emission, $FWHM=1$\,km\,s$^{-1}$, and optical depth
consistent with the assumed $T_{\rm ex}$ value and the maximum antenna
temperature of the spectrum. Figure~\ref{fig:n2h+_spectra} shows examples of
fitted spectra.

The runtimes in the analysis of the $R=1-32$ maps are included in
Fig.~\ref{fig:timings}. The speed on the GPU is about $10^5$ spectra per
second, a few times slower than for the single-component Gaussians fits. The
increase in the run time is partly due to the larger number of free parameters
(although still fewer than for two-component Gaussian fits) and some small
overhead from the use of the penalty functions. However, the main cause is
simply the larger number of velocity channels, 224 channels compared to the
120 channels in the previous Gaussian fits.

\subsection{Test with synthetic spectra with  Gaussian components}
\label{sect:test_gaussian}

In this section we study simple synthetic spectra with three Gaussian velocity
components. These are used to examine the absolute accuracy of the parameter
fits as well as the selection of the number of velocity components. The model
selection was done using the Akaike information criterion (AIC), which was
calculated as $AIC=2 k - 2 \ln L$, where $k$ is the number of free parameters
in the model and $L$ is the maximised likelihood.\footnote{We omit the
correction for small samples, which would change the expression to $AIC = 2 k
- 2\ln L + \frac{2 k^2+2 k}{n-k-1}$. The corrected formula would set more
preference for models with fewer parameters. However, the term is small,
$\sim0.1$ between our consecutive Gaussian models. To our knowledge, the
correction term is also proven only for linear models \citep{Cavanaugh1997}.}
AIC is but one possible criterion that can be used to choose the optimal
number of fitted components, based on the completed fits.

\begin{figure}
\includegraphics[width=7.8cm]{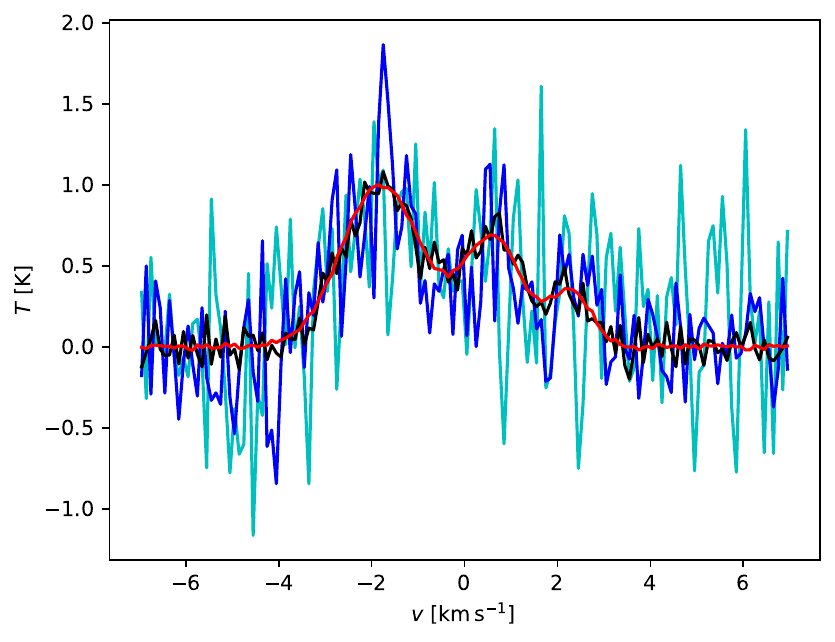}
\caption{
Examples of synthetic spectra with three velocity components. The red, black,
blue, and cyan lines correspond to the respective noise levels of 1\%, 10\%,
30\%, and 50\% of the peak value in the noiseless spectrum.
}
\label{fig:simple_3g_spe}
\end{figure}

The fitted spectra contain Gaussian components where the peak intensities are
1.0, 0.67, and 0.33\,K, central velocities -1.8, 0.6, and 2.3\,km\,s$^{-1}$,
and $FWHM$ values 2.0, 1.5, and 0.5\,km\,s$^{-1}$, for the three components,
respectively. The noise was varied in 50 steps between 1\% and 50\% of the peak
intensity, before adding this normal-distributed noise.  We fitted 1000 noise
realisations for each noise level to estimate the distribution of the
recovered parameters. Figure~\ref{fig:simple_3g_spe} shows examples of noise
realisations, where the three peaks are clearly visible at low noise levels.

The selection of initial values is an important step of model fitting. Apart
from reading initial values from external files, SPIF provides a few ways to
set the initial values (cf. Sect.~\ref{sect:implementation} and
Appendix~\ref{app:manual}). For the first velocity component, we used the
intensity and velocity of the spectrum maximum as the initial values. For the
second and third component, the intensities were initialised to the same
values (i.e. overestimating the true values). The velocities were set to fixed
values of 0 and 1\,km\,s$^{-1}$, respectively, and the $FWHM$ to
1\,\,km\,s$^{-1}$. Thus, apart from the intensity and velocity of the first
Gaussian, the selected initial values are not particularly close to the true
values. This is compensated by repeating the fits with more than twenty times
with randomised initial values. The random shifts were generated from $N(0,
\sigma)$, with a standard deviation of $\sigma=0.2$ units (K or km\,s$^{-1}$).
Thus, the range of initial values does not yet directly cover, for example,
the actual radial velocity of the weakest velocity components.

\begin{figure*}
\sidecaption
\includegraphics[width=12cm]{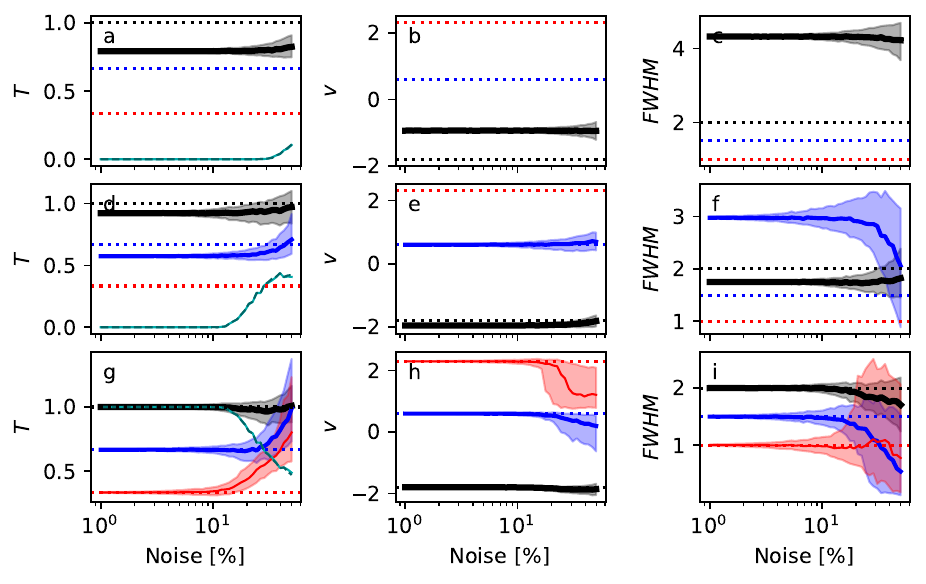}
\caption{
Accuracy of fit parameters in the case of spectra with three Gaussian
components. The three rows of frames correspond to the one to three component fits. The
solid black, blue, and red lines show the median of the parameter estimates as
a function of the noise level, and the shaded areas of the same colour show
the inter-quartile ranges. The dotted horizontal lines correspond to the true
values. In the frames a, d, and g, the light cyan lines show the fraction of
cases where AIC preferred fits with the corresponding number components (i.e.
one, two, or three). The dashed dark cyan lines are the same after 40 channels
without significant emission were excluded, but these overlap almost perfectly
with the previous cyan lines.
}
\label{fig:simple_3g}
\end{figure*}

Figure~\ref{fig:simple_3g} shows the parameters recovered by the 1-3 component
Gaussian fits. The single-component fit tries to match all emission, resulting
in a mean velocity between the strongest emission components and a large
estimated $FWHM$ of $\sim$4\,km\,s$^{-1}$. With two fitted Gaussians, the
first one already gives a good approximation of the strongest emission
component while the second fitted Gaussian approximates the sum of the
remaining two. When the model uses three Gaussians, these match the three real
components accurately up to $\sim$10\% noise level. Thereafter the intensity
of the two weaker features gets systematically overestimated (with S/N$<$3 for
the weakest one), and they move in velocity towards the strongest component. 

Using all the velocity channels shown in Figure~\ref{fig:simple_3g}, the AIC
criterion clearly prefers the three-component model for all noise levels below
$\sim$10\%. Thereafter, as the accuracy of the parameters of the weakest
component starts to degrade, also the AIC criterion sometimes prefers the
two-component model. However, even at the 50\% noise level, the
three-component model would still be selected in about half of the cases. 

We also checked, how sensitive the AIC criterion is on the inclusion of
channels without significant emission. We repeated the analysis after removing
20 channels from both ends of the spectra and thus reducing the total number
of channels by some 30\%. Although this does change the AIC values, the effect
on the model selection is negligible, as shown in Fig.~\ref{fig:simple_3g}.
This is not entirely surprising. One can add any number of noisy channels in
the AIC calculation, and, as long as the model prediction in these channels is
zero, all AIC values increase by the same constant, with no change in their
magnitude order.  If the model used one component to fit a pure noise feature,
that should still get penalised in the AIC comparison. Alternatively, that can
be prevented more directly by removing such channels from the analysis
altogether, if the channels are known to contain no signal.

\subsection{Tests with synthetic spectra from the MHD simulation} \label{sect:precision}

In this section we examine the precision to which the fitted components
reproduce selected properties of the observed spectra. This concerns the
differences between complex realistic spectra and the simplified
spectral models used in the fits. The accuracy of the error estimates of
the fitted parameters is examined later in Sect.~\ref{sect:error_estimates}.
Here we look only at direct observables, such as the line areas and the
velocity dispersion. The more complex question of the connections between the
fit parameters and the actual physical source properties (such as the true
column density) is mostly beyond the scope of the present paper.

\subsubsection{Line area in Gaussian fits} \label{sect:line_area}

One of the main parameters extracted from Gaussian fits is the line area,
which for optically thin lines would also be proportional to the column
density (apart from the effects of the line-of-sight $T_{\rm ex}$ variations).
It is therefore important to know, how accurately the Gaussian fits are likely
to approximate the complex emission from an interstellar cloud, such as
approximated by the synthetic observations of the MHD cloud model.

\begin{figure}
\sidecaption
\includegraphics[width=8.8cm]{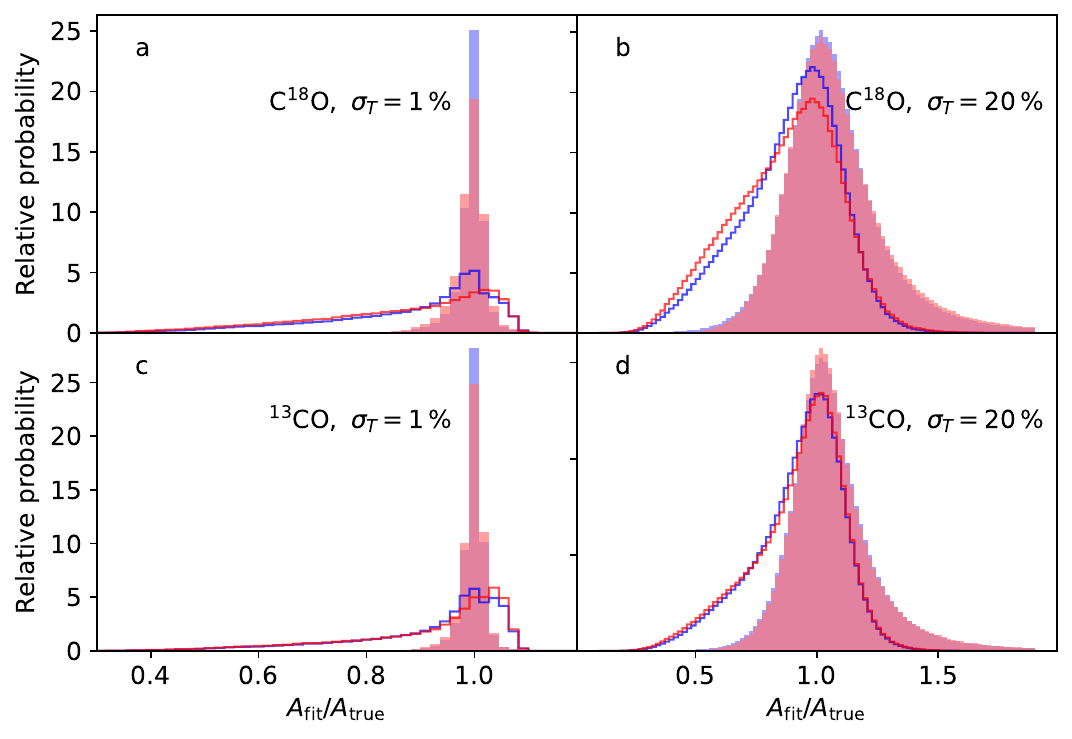}
\caption{
Line areas from Gaussian fits versus the true line area. Results are shown for
C$^{18}$O and $^{13}$CO (upper and lower frames, respectively) and for noise
levels of $\sigma_{T}=1$\% and $\sigma_{T}=20$\% of the peak value in each
spectrum (left and right frames, respectively). The blue histograms contain
all spectra with $T_{\rm}^{\rm max}>10$\,mK and the red histograms the
strongest spectra with $T_{\rm}^{\rm max}>1$\,K. The unfilled histograms
correspond to single-component fits and the filled histograms to one to three
component fits, where the number of components was selected based on the AIC
criterion.
}
\label{fig:areas}
\end{figure}

We examined the ratio between the line area provided by Gaussian fits,
performed at different noise levels, and the true line areas that were
obtained by directly summing the channel values in the original noiseless
spectra. Figure~\ref{fig:areas} shows the results for C$^{18}$O and $^{13}$CO
spectra with 1\% or 10\% of noise (relative to the maximum value in each
spectrum). The line areas of the Gaussians are shown for single-component fits
(unfilled histograms) and for 1-3 component fits, where the number of
components is selected based on AIC. 
The calculation is in this case done using all channels (a velocity range of
12\,km\,s$^{-1}$), although the emission often covers less than half of the
full bandwidth (cf. Fig.~\ref{fig:spectra}).
At the $\sigma_T=1$\,\% noise level, the best Gaussian model recovers the line
area relatively accurately, with an rms error of a few percent. If one uses
only a single Gaussian, the distribution peaks close to one but, as expected,
has a long tail towards lower values, due to the other emission in the
spectra. When the noise is increased to 10\%, the differences between the
different fits decrease. Although the noise increases in the multi-component
fits, they remain nearly perfectly unbiased. For the 10\% noise level, the
Gaussian fits recover the total line area usually with better than $\sim$20\%
accuracy. However, the accuracy will be worse for complex spectra, and errors
larger than 50\% are encountered in $\sim$1\% of the spectra.

Figure~\ref{fig:areas} shows no significant differences between C$^{18}$O and
$^{13}$CO. This is mainly because the S/N was set equal for every spectrum, and
the fraction of optically thick $^{13}$CO spectra is still small.  In actual
observations, the differences between the lines would arise mostly from the
different S/N. Our results show no clear difference between the spectrum
samples where the peak intensities are $T_{\rm A}^{\rm max}$>10\,mK or $T_{\rm
A}^{\rm max}$>1\,K, in spite of the fact that stronger lines are seen
preferentially towards dense regions that are contracting gravitationally and
are more likely to have multiple line-of-sight components.

\subsubsection{Velocity dispersion from Gaussian fits} \label{sect:velocity_dispersion}

\begin{figure}
\sidecaption
\includegraphics[width=8.8cm]{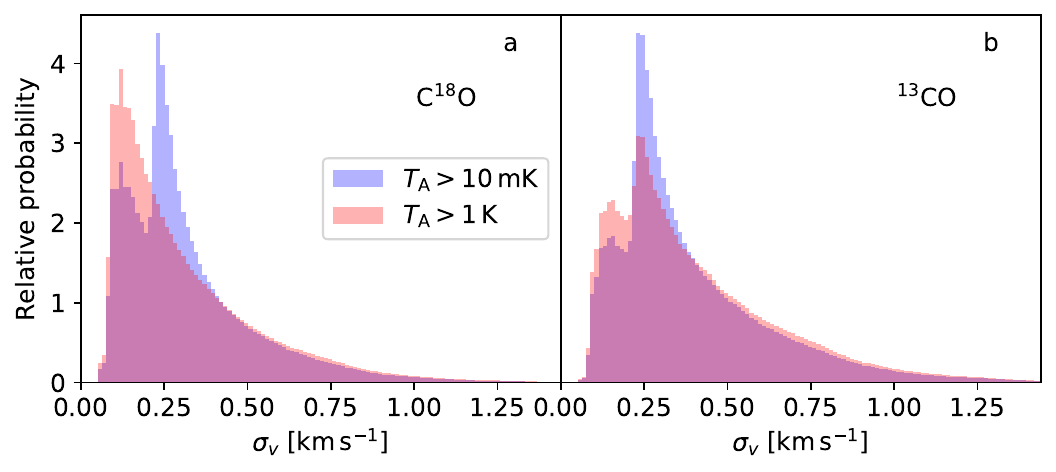}
\caption{
Distribution of the velocity dispersion $\sigma_{v}$ of the synthetic
C$^{18}$O and $^{13}$CO spectra without added observational noise. The
histograms correspond to spectra with $T_{\rm}^{\rm max}>10$\,mK (blue
histograms) and $T_{\rm}^{\rm max}>1$\,K (red histograms). 
}
\label{fig:disp0}
\end{figure}

We next look at the velocity dispersion deduced from Gaussian fits. Before
examining the fits, Fig.~\ref{fig:disp0} shows the overall statistics of the
velocity dispersion in the synthetic noiseless spectra. The $\sigma_{v}$
distribution peaks at low values$\sim$0.3\,km\,s$^{-1}$ and moves to lower
values for brighter spectra. The values are bound from below by the thermal
linewidth (e.g. $\sim$0.05\,km\,s$^{-1}$ at $T_{\rm kin}$=10\,K), but the
distributions have a long tail that extends even beyond the range shown in
Fig.~\ref{fig:disp0} (not visible on the linear scale used in the plot).

\begin{figure}
\sidecaption
\includegraphics[width=8.8cm]{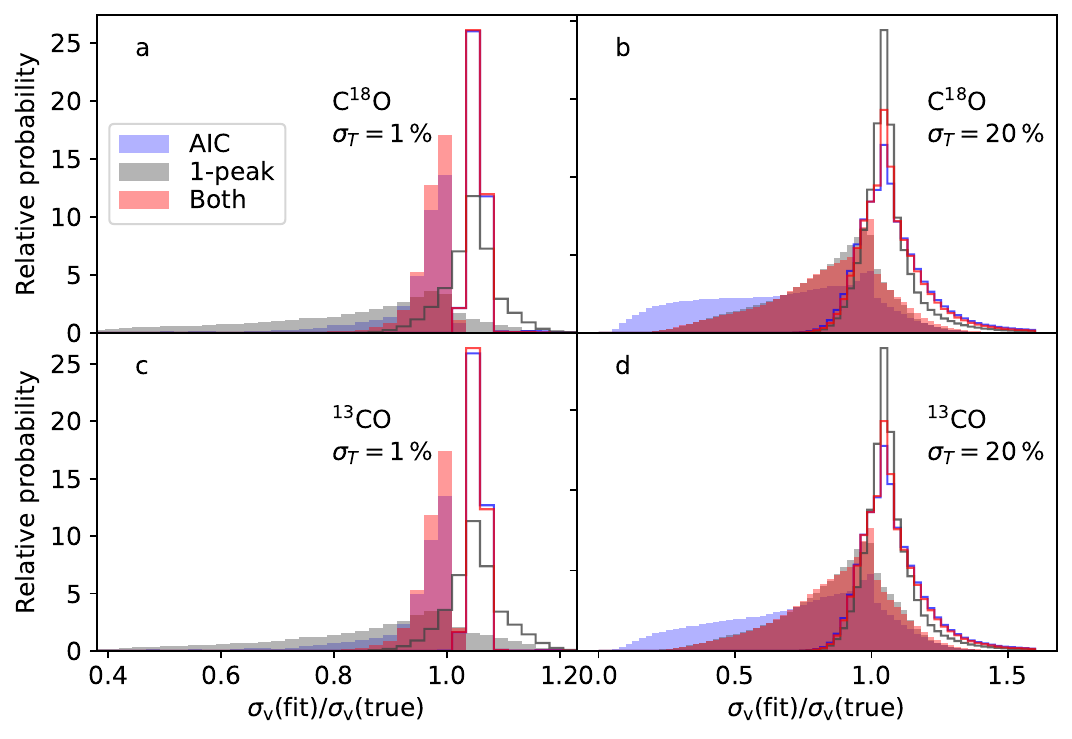}
\caption{
Histograms for the ratio of the velocity dispersion recovered by
single-component Gaussian fits ($\sigma_{v}({\rm fit})$) and the true velocity
dispersion of the synthetic spectra ($\sigma_{v}({\rm true})$).  Histograms
correspond to spectrum subsets where the best model has only a single velocity
component (AIC, based on the AIC criterion), spectra have only a single
peak (``1-peak''), and the intersection of these two samples (``both''). The
true velocity dispersion was calculated either for the whole spectrum (filled
histograms) or only within $2.5\sigma_{v}(\rm fit)$ around the fitted centre
velocity (unfilled histograms). Results are shown for C$^{18}$O and $^{13}$CO
(upper and lower frames, respectively) and for noise levels of $\sigma_T$=1\%
and $\sigma_T$=20\% (left and right frames, respectively).  
}
\label{fig:dispersion}
\end{figure}

Figure~\ref{fig:dispersion} compares the standard deviation of the
fitted Gaussian components $\sigma_{\rm v}({\rm fit})$ to the actual
velocity dispersion $\sigma_{\rm v}({\rm true})$ of the synthetic spectra. The
spectra correspond to the $R=2$ case and all the three view directions,
excluding spectra with $T_{\rm}^{\rm max}<1$\,mK. The comparison is further
limited to samples where a single Gaussian component could be expected to give
a good representation of the emission. The first sample includes spectra where
the single-component fit is preferred based on the AIC criterion. The
distribution of the ratio $\sigma_{v}({\rm fit})$/$\sigma_{v}({\rm true})$
shows two shallow peaks, one around one and another extending down below the
ratio of 0.5. This is not surprising, since the spectra often have emission
extending over a velocity range several times larger than the individual
narrow components. The second sample includes only truly single-peaked
spectra, as determined from the noiseless spectra (no multiple local maxima
separated by a dip of more than 10\%, cf. Appendix~\ref{app:stat}). This
narrows the distribution only slightly, as almost all of these spectra also
fulfil the AIC criterion for a single component. The sample contains still
about one million spectra, some 60\% of the spectra in the $R=2$ maps. It is
only when the calculation of the reference (``true'') velocity dispersion is
limited to channels within $2.5 \sigma_{v}({\rm fit})$ of the fitted centre
velocity that the $\sigma_{v}({\rm fit})$/$\sigma_{v}({\rm true})$
distribution peaks more clearly close to the value of one. In that case the
fitted Gaussians tend to even slightly overestimate the velocity dispersion of
the $\pm 2.5 \sigma_{v}({\rm fit})$ velocity interval. The distribution is
still non-Gaussian, with tails extending close to $\sim$20\%, even in the case
of low noise ($\sigma_T$=1\%). At the higher noise level, the distributions
are much wider and the overestimation can reach values above 50\%. Although
the $^{13}$CO spectra are somewhat wider and sometimes optically thick, there
is again no clear difference between the C$^{18}$O and $^{13}$CO
distributions.

Figure~\ref{fig:dispersion} was restricted to cases where the spectra appeared
to have just one component, and it showed how difficult it is to estimate the
velocity dispersion of a single component. This is even more true if the
component needs to be separated from other significant emission features
\citep[e.g.][]{Hacar2018,Lu2022}. Based on Fig.~\ref{fig:dispersion}, the
uncertainty can easily reach 20\% in the velocity dispersion and,
consequently, would be $\sim$50\% in the estimated kinetic energy. If the S/N
of the spectrum is sufficient, it may also be better to estimate the velocity
dispersion directly from the observed spectrum, instead of relying on Gaussian
fits of non-Gaussian profiles.

\subsubsection{Parameters of hyperfine structure lines} \label{sect:hfs_precision}

\begin{figure}
\sidecaption
\includegraphics[width=8.8cm]{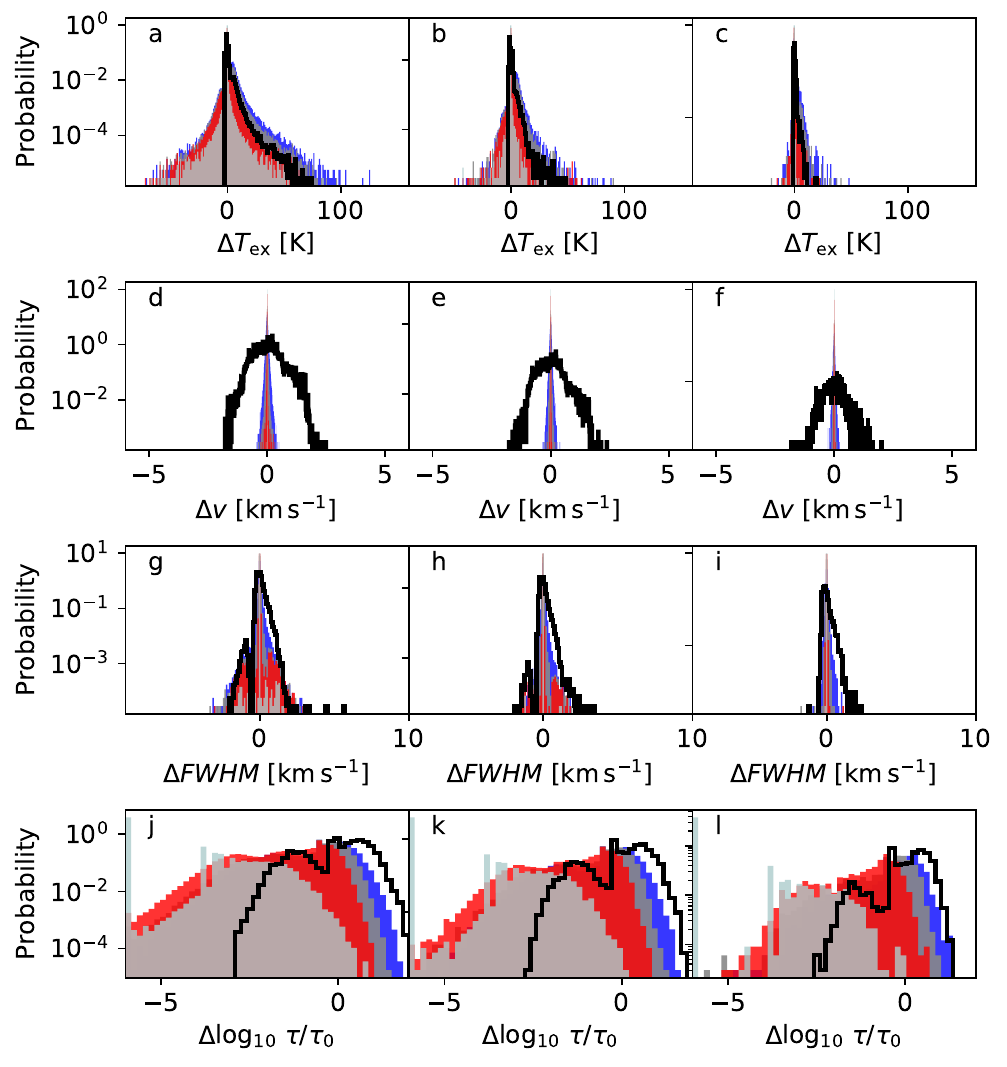}
\caption{
Dispersion of parameter estimates in ${\rm N_2H^{+}}$ fits of synthetic
spectra from the MHD model.  From left to
right, the frames correspond to synthetic observations in the $R$=2, 8, and 32
maps. The filled histograms (in order of increasing width) correspond to noise
levels of 1\%, 3\%, 10\%, and 20\%. They correspond to the parameter estimates
relative to those obtained for spectra without noise (absolute difference
on first three rows and logarithm of the ratio in the bottom row). The latter
are plotted for reference with solid black lines and shifted to the mean value
of zero.
}
\label{fig:hfs_stat}
\end{figure}

Figure~\ref{fig:hfs_stat} illustrates the precision of the parameter values
obtained from $\rm N_2 H^{+}$(1-0) spectra. Unlike in Gaussian fits, the true
values of the fitted parameters are not known in a similarly straightforward
manner. Therefore, Fig.~\ref{fig:hfs_stat} examines the estimates relative to
those obtained in fits of noiseless spectra, which are also used to illustrate
the overall range of parameter values. Each histogram in
Fig.~\ref{fig:hfs_stat} contains data from all three view directions,
excluding spectra with $T_{\rm}^{\rm max}<$10\,mK. When the sample is further
reduced to those that are single-peaked in C$^{18}$O (one velocity
component), one is left with 345~340, 24~486, and 2041 spectra for $R$=2, 8,
and 32, respectively.

The values of velocity and $FWHM$ are recovered very accurately, and even with
20\% of added noise, the radial velocity is accuracy to within
$\sim$0.1\,km\,s$^{-1}$. The $FWHM$ is less precise than the velocity itself,
and the different relative to the analysis of noiseless spectra can reach even
2\,km\,s$^{-1}$. If one had not removed the spectra with multiple velocity
components (the larger sample having $\sim$1.49 million, 94~603, and 5869
spectra for $R$=2, 8, and 32, respectively), the $v$ and $FWHM$ distribution
would be wider by a factor of at least a factor of $\sim$3.

The $T_{\rm ex}$ and $\tau$ estimates are affected by strong parameter
correlations, and they show much wider scatter (i.e. dependence on the noise
level and the noise realisation). In Fig.~\ref{fig:hfs_stat}j-l, the $\tau$
estimates often vary by more than 100\% ($\Delta \,\log_{\rm 10} \tau>0$).
With the adopted antenna temperature limit of 10\,mK, the sample contains
optically thin spectra, for which $T_{\rm ex}$ and $\tau$ are degenerate and
an individual parameters is almost completely unconstrained. In real
observations, many of those spectra would be excluded because of the practical
limitations of noise.

\begin{figure}
\sidecaption
\includegraphics[width=8.8cm]{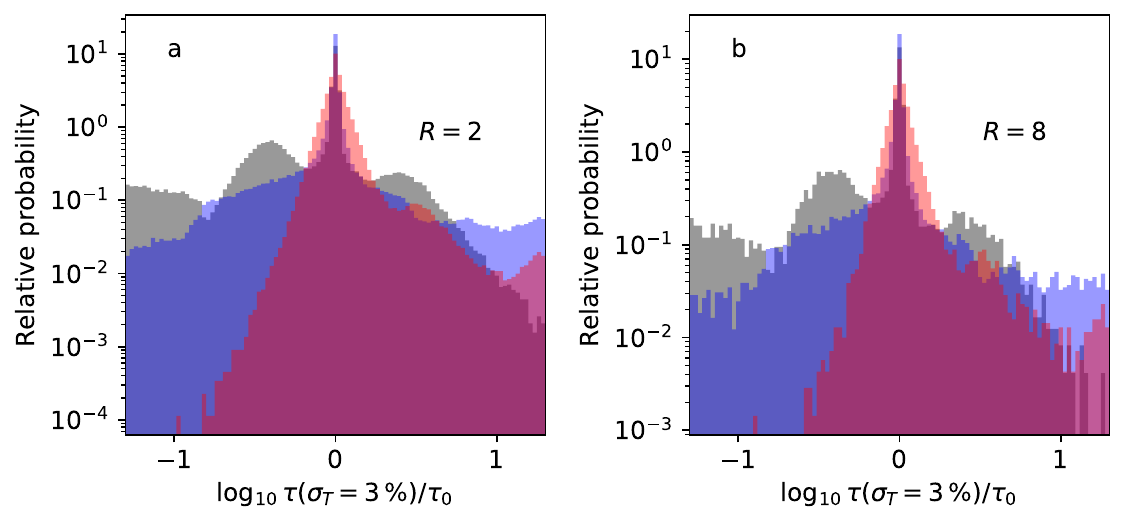}
\caption{
Histograms for the ratio of optical depths derived from fits to noiseless and
noisy N$_2$H$^{+}$ spectra from the MHD model. The colours correspond to
optical depth ranges of $\tau=0.005-0.05$ (grey), $\tau=0.05-1$ (blue), and
$\tau>1$. The observational noise is 3\% of the peak antenna temperature
value, and the observations correspond to the $R=2$ (frame a) and $R=8$ (frame
b) spatial resolutions.
}
\label{fig:hfs_stat2}
\end{figure}

\begin{figure}
\sidecaption
\includegraphics[width=8.8cm]{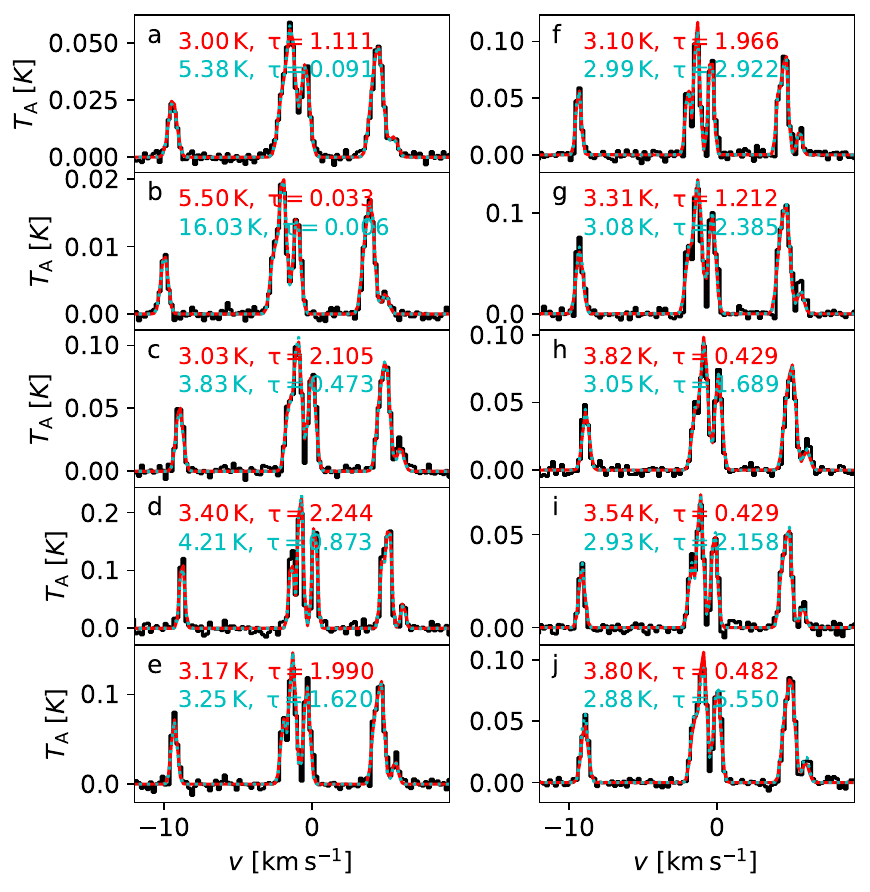}
\caption{
Examples of N$_2$H$^{+}$ spectra from the MHD model where the ratio of
optical depths derived from noiseless and noisy spectra ranges from $\sim$0.1
(frame a) to $\sim$10 (frame j). The black histograms show the observed
spectra that contain 3\% relative noise. The fits to noiseless and noisy
spectra are plotted with dashed red and dotted cyan lines, respectively. The
$T_{\rm ex}$ and $\tau$ estimates from the fits are listed in the frames in
the same colours.
}
\label{fig:hfs_stat3}
\end{figure}

Figure~\ref{fig:hfs_stat2} separates the previous optical-depth estimates
based on the values of the estimated $\tau$, for $\rm N_2 H^{+}$ spectra
at 3\% noise level. The estimates are peaked strongly around ratio one, the
optical depth being the same in the analysis of noiseless and noisy spectra.
However, there is a fraction of spectra (a factor of 10$^2$-10$^3$ below the
mode of the distributions) where the estimates differ, and the probability for
deviations of one order of magnitude are almost as common as smaller changes.
Changes larger than a factor of two are observed for 30\% of the optically
thin spectra ($\tau=0.005-0.05$), the fraction dropping to 5\% for spectra
with $\tau>1$. However, we note than a total optical depth $\tau=1$
corresponds to an optical depth of only 0.26 for the main hyperfine
component. Large changes in the fitted parameters could point to an inaccuracy
in the fits. However, Fig.~\ref{fig:hfs_stat3} shows ten sample spectra where
the ratio of the optical depths derived from spectra without and with noise
varies from $\sim$0.1 to 10. In spite of the large differences in $\tau$, all
fits are consistent with the data. As the synthetic spectra are the result of
a complex line-of-sight integral of emission from different radial velocities
and excitation temperatures, they never precisely match the fitted spectral
model, and this may also increase the sensitivity to the noise.

\begin{figure}
\sidecaption
\includegraphics[width=8.8cm]{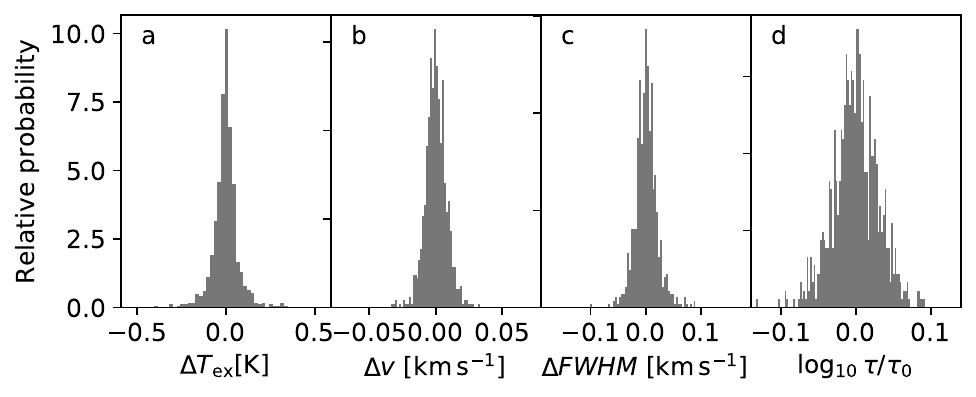}
\caption{
Comparison of N$_2$H$^{+}$ hyperfine parameters in fits to noiseless
spectra and to spectra with 3\% relative noise. The sample includes only the
lines of sight for 10\% of the highest column densities in the MHD model.
}
\label{fig:2-hfs_colden}
\end{figure}

Since the uncertainty of $T_{\rm ex}$ and $\tau$ values is linked to the low
optical depths, we examined separately the spectra that correspond to 10\% of
the highest column densities in the MHD model. The column densities are
obtained directly from the 3D model and thus represent high true optical
depths (instead of just high estimated optical depths). The selected spectra
also correspond more closely to the regions where N$_2$H$^{+}$ emission would
be in practice observable. In this sample, the average peak temperature is
approaching 5\,K, and the average optical depth of the main component is
$\sim$3. The results are shown in Fig.~\ref{fig:2-hfs_colden}. When compared
to the statistics of all the spectra shown Fig.~\ref{fig:hfs_stat2} (the
middle column with $R=8$), the much higher accuracy of the $T_{\rm ex}$ and
$\tau$ estimates is evident.

\subsection{Error estimates} \label{sect:error_estimates}

In SPIF, the main method of error estimation is Monte Carlo simulation. The
accuracy of the error estimates can be checked with simulations, However, the
simulations are nearly equal to the Monte Carlo estimation itself, with the
exception that in the test we can simulate random noise realisations based on
true spectra, while the Monte Carlo noise estimation starts with
observations that already contain noise.

\subsubsection{Gaussian fits}

We look first at simple synthetic spectra that contain one Gaussian component
and normal-distributed noise with $\sigma(T_{\rm A})$ ranging from 1\% to 30\%
of the peak value of the spectrum. We examine 322 noise values, each with 322
independent noise realisation, giving slightly more than 100 000 spectra in
total. 

\begin{figure}
\sidecaption
\includegraphics[width=8.8cm]{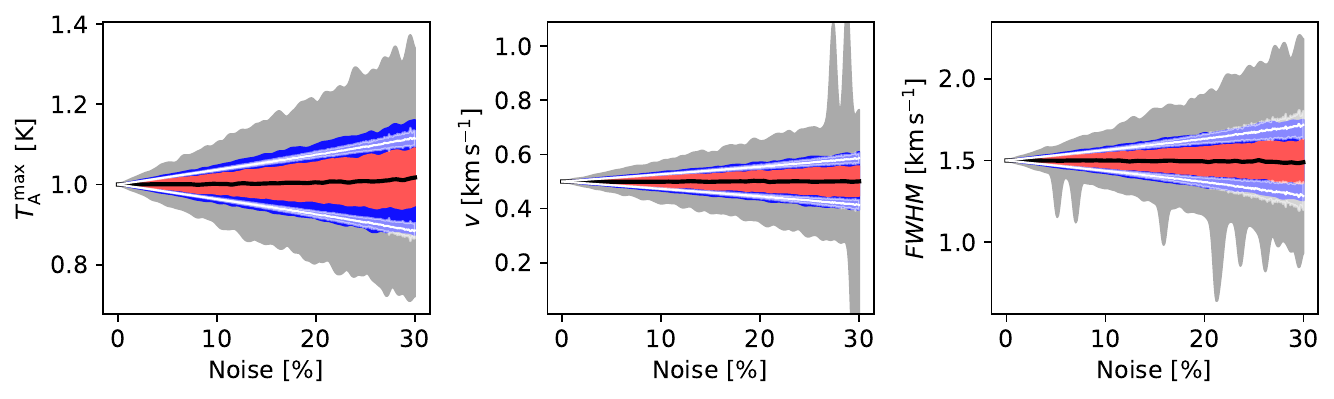}
\caption{
Parameter uncertainties in single-component Gaussian fits to spectra
containing one Gaussian. The shaded regions show the error distributions for
the estimated parameters as a function of the relative noise 1-30\%. The grey,
blue, and red colours correspond to [0.01, 99.9], [10, 90], and [25, 75]
percentile intervals, respectively. The solid black line shows the median
estimates. The solid white lines correspond to the mean 1-$\sigma$ error
estimates (plotted relative to the true values), as obtained from 100 Monte
Carlo samples. The white shaded regions correspond to the 1-$\sigma$
dispersion of the error estimates between spectra at the same noise level.
}
\label{fig:4a}
\end{figure}

Figure~\ref{fig:4a} shows the resulting scatter in the parameter estimates and
the error estimates derived for each spectrum with 100 Monte Carlo samples.
The small sample of 100 already gives an accurate picture of the $1-\sigma$
parameter uncertainties, and even the 1-$\sigma$ dispersion of the error
estimates themselves (white shaded regions in Fig.~\ref{fig:4a}) is relatively
small compared to the error estimates. However, if one wants to characterise
the tails of the error distribution, beyond the 1-$\sigma$ level, the
computational cost increases rapidly, inversely proportionally to the
probability contained in those tails.

\begin{figure}
\sidecaption
\includegraphics[width=8.8cm]{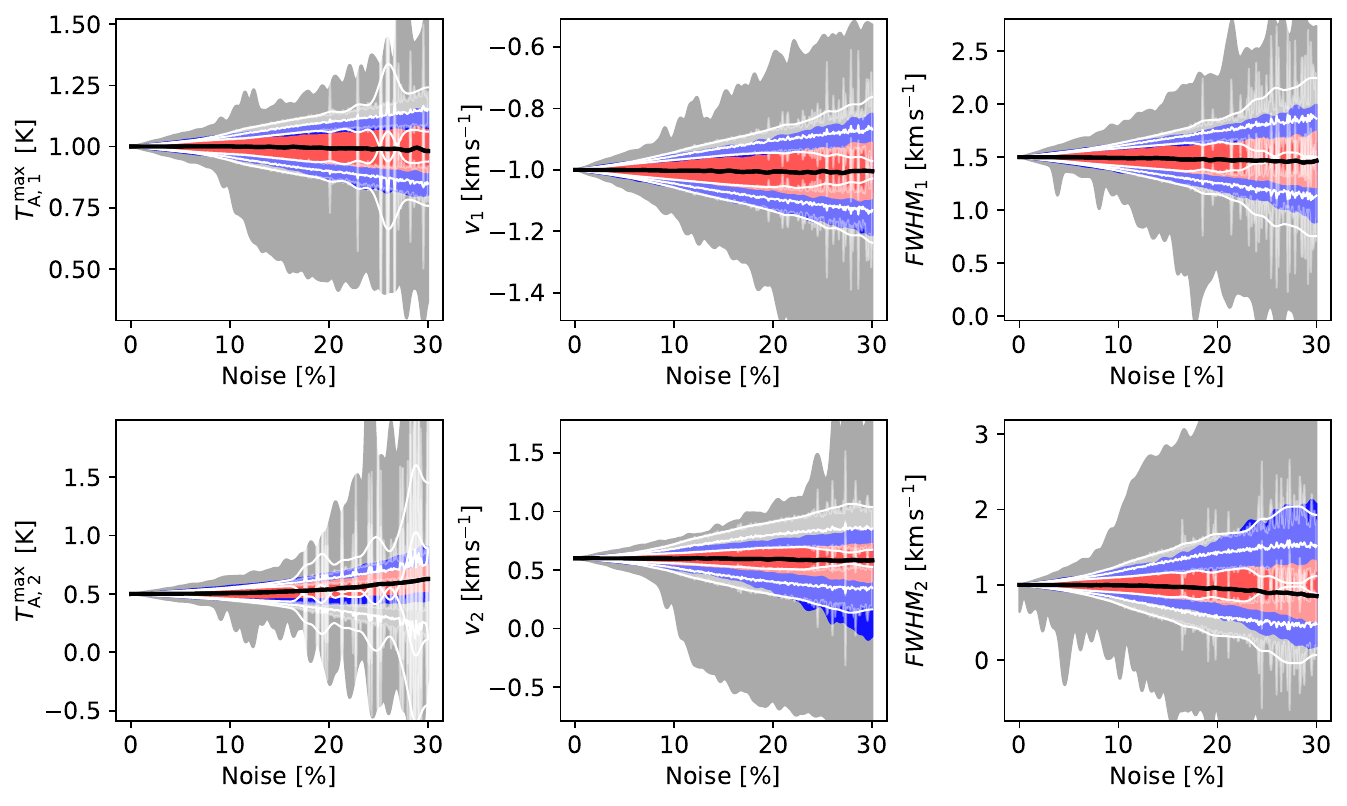}
\caption{
As Fig.~\ref{fig:4a} but for two partially overlapping Gaussian components.
}
\label{fig:4b}
\end{figure}

We tested next the fitting of two Gaussians that had peak intensities of 1\,K
and 0.5\,K, central velocities of -1\,km\,s$^{-1}$ and 0.5\,km\,s$^{-1}$, and
$FWHM$ values of 1.5\,km\,s$^{-1}$ and 1.0\,km\,s$^{-1}$, respectively.
According to Fig.~\ref{fig:4b}, the results are mostly unbiased, although for
the weaker component the $T_{\rm A}$ estimates increase and the $FWHM$
estimates decrease with increasing noise. The errors are larger than in the
one-component case, and the distribution of radial velocity errors shows
stronger tails. For each Monte Carlo sample, the lower of the two radial
velocities is always assigned to the first component, which results in some
asymmetry in the corresponding error distributions. The error estimates are on
average correct, although, compared to Fig.~\ref{fig:4a}, more noisy. In
Fig.~\ref{fig:4b} the 1-$\sigma$ limits of the error estimates have been
smoothed (Gaussian averaging, with FWHM equal to three steps in noise).
Nevertheless, they still occasionally show fluctuations exceeding a factor of
two (i.e. a large difference between realisations of similar noise). This can
be due to pure noise in the error estimates, but is partly true variation:
some noise realisations lead to larger uncertainty in the parameter estimates.
For these data, AIC prefers the two-component model over one-component model,
with very few exceptions at the highest noise levels.

\begin{figure*}
\sidecaption
\includegraphics[width=12cm]{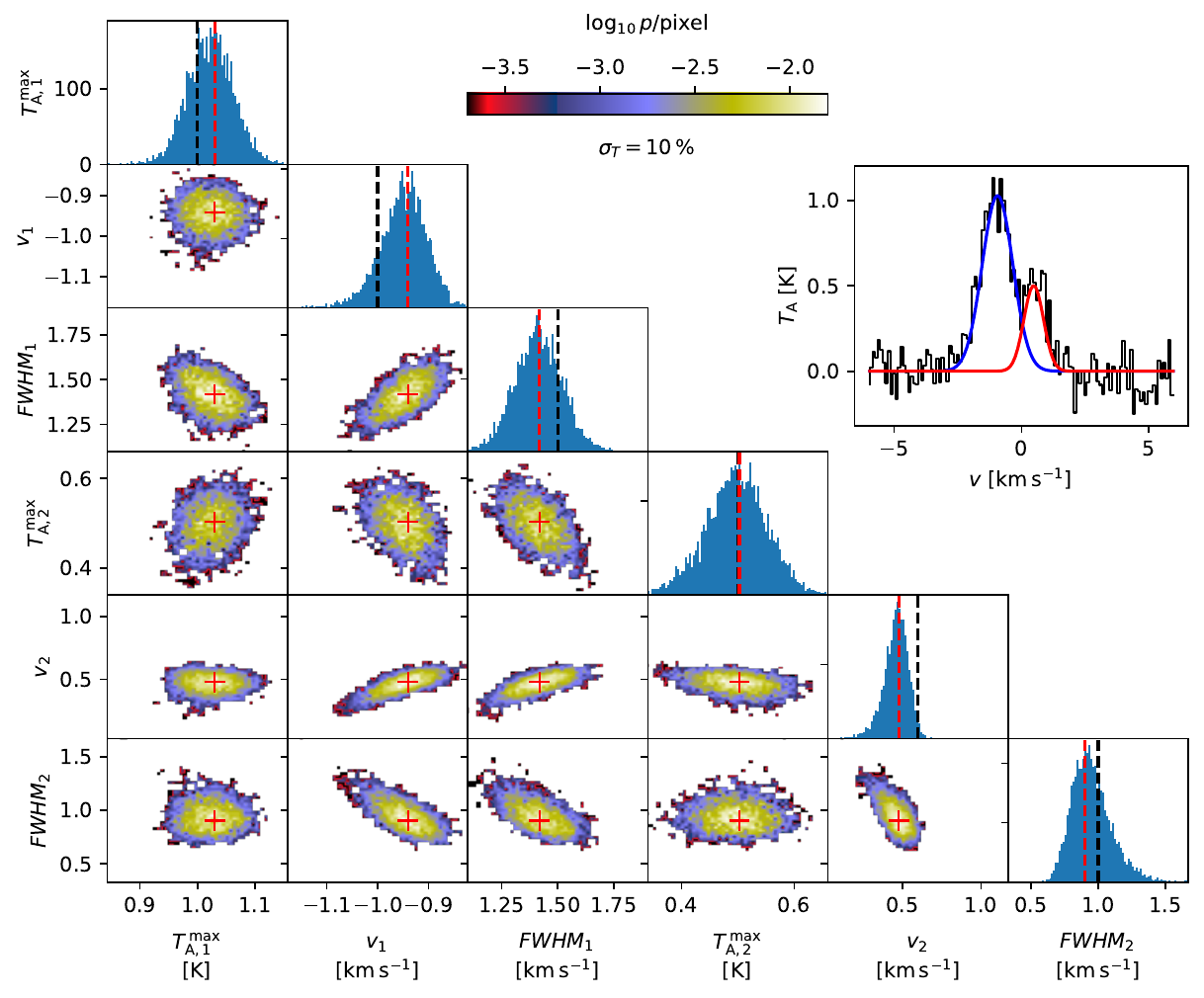}
\caption{
MCMC error estimates for a two-component Gaussian fit. The synthetic spectrum
consists of two Gaussian components and white noise with $\sigma_{T}$=10\%
relative to the peak value in the spectrum. The corner plot shows the
parameter correlations, where the colours of the 2D histogram correspond to
the logarithm of probability (MCMC samples per pixel), as indicated by the
colour bar on top. The red crosses correspond to the least-squares
solution. The diagonal frames show the histograms for the individual
parameters, and the dashed black and red lines correspond to the true values
(before observational noise is added) and to the least squares fit,
respectively. The frame on the right shows this particular fitted spectrum
(black line) and the two components corresponding to the $\chi^2$ minimum.
}
\label{fig:mcmc_2g}
\end{figure*}

Although the MCMC procedure is less robust than the direct Monte Carlo
simulations above, reasonable for MCMC error estimates could be calculates for
both the single-component and two-component cases. We reduced the sample of
spectra to just 30 realisation at each noise level (a total of 10304 spectra),
mainly to reduce storage requirements ($\sim$1\,GB for each set of MCMC
samples) rather than the runtimes. We started the MCMC chains with the maximum
likelihood solution and ran a burn-in phase of 1000 steps, during which the
step sizes was adjusted to reach $\sim$30\% acceptance ratio. We thereafter
registered for each spectrum 5000 samples (of the 3 or 6 free parameters),
from every 100th MCMC step. Each individual MCMC step is much faster than the
full $\chi^2$ optimisation, which means that calculations were not more
time consuming than the previous runs for the Monte Carlo error estimates. The
total runtime of the two-component MCMC calculations was about 13 seconds
(including the host code and the reading and writing of files for the 10304
spectra), of which some 10 seconds was spent on actual computations within the
OpenCL kernels. This corresponds to an average speed of about 1000 spectra per
second. Figure~\ref{fig:mcmc_2g} shows the results for one spectrum with 10\%
relative noise.

\begin{figure}
\sidecaption
\includegraphics[width=8.8cm]{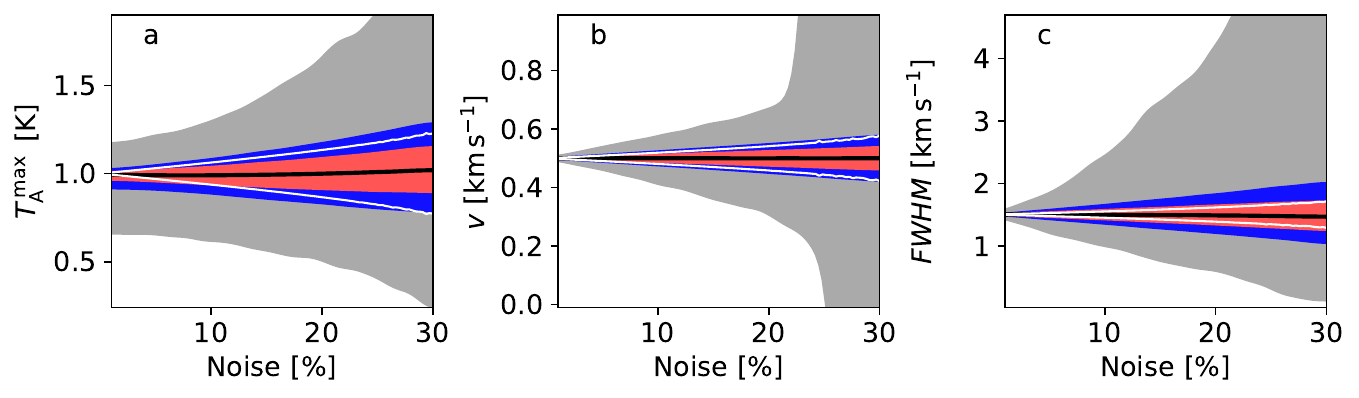}
\caption{
As Fig.~\ref{fig:4a} but for single-peaked synthetic C$^{18}$O spectra
selected from the MHD cloud model. The white lines show the average
$\pm$1-$\sigma$ Monte Carlo error estimates.
}
\label{fig:4c}
\end{figure}

Apart from the noise, in the previous examples the spectra matched perfectly
the models that were fitted to them. For comparison, we examined the synthetic
C$^{18}$O observations of the MHD model, concentrating on those single-peaked
$R=2$ spectra where the emission rises above 10\,mK. We took a random
selection of 10 000 such spectra, added 1-30\% of observational noise (at 100
discrete noise levels), registered the maximum likelihood parameter estimates,
and estimated their errors based on 100 Monte Carlo samples. The goal was to
see if the uncertainties are comparable to those in Fig.~\ref{fig:4a} or
whether other emission in the spectra and deviations from the perfect Gaussian
line shapes has a noticeable effect on the errors. In Fig.~\ref{fig:4c}, we
rescale the results to a peak value of 1\,K, move the central velocity to
0.5\,km\,s$^{-1}$, and scale the $FWHM$ estimates to 1.5\,km\,s$^{-1}$ in
order to enable more direct comparison to Fig.~\ref{fig:4a}.  The errors are
larger but this is visible mostly in the tails of the error distribution (in
the [0.01,99.9] percentile intervals) and at the higher noise levels (above
$\sim$20\% relative noise). The line-of-sight confusion of course varies from
spectrum to spectrum, and the largest uncertainties exceed even those seen in
the two-component fits of Fig.~\ref{fig:4b}.

\subsubsection{Hyperfine structure lines} \label{sect:hyperfine2}

Section~\ref{sect:hfs_precision} showed that in hyperfine fits the parameters
$T_{\rm ex}$ and $\tau$ are often individually not well constrained. This
should be reflected in the error estimates, but it can also complicate the
error estimation itself. We examined simulations, where the spectra contained
only one velocity component, the emission followed exactly the hyperfine
model, and the noise was varied between 1\% and 20\% of the maximum antenna
temperature. The excitation temperature was set to $T_{\rm ex}$=9\,K and the
total optical depth to $\tau$=0.5. To facilitate comparison with some results
obtained with routines from the Scipy library, the SPIF fits included 322
noise levels with 100 spectra each, but only $322 \times 5$ spectra were used
in the comparison to the Scipy fits.

\begin{figure}
\sidecaption
\includegraphics[width=8.8cm]{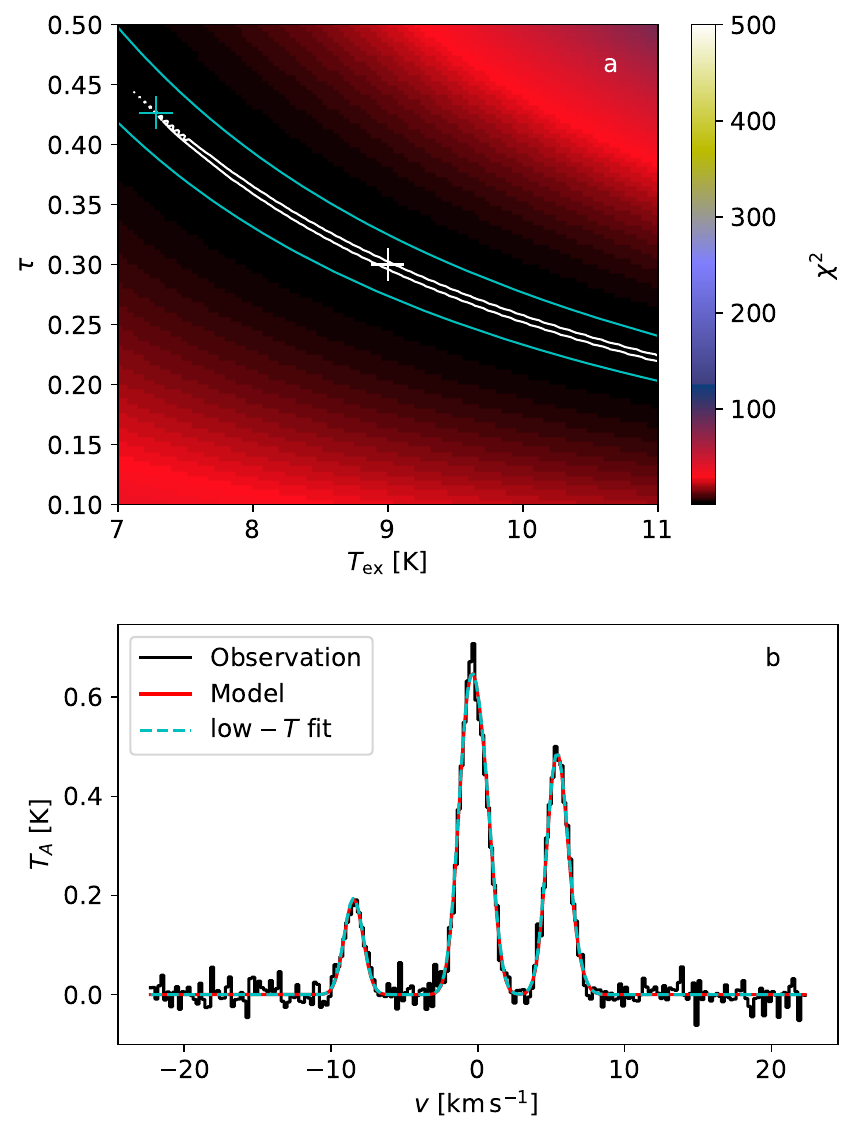}
\caption{
Example of hyperfine fits with degeneracy between the $T_{\rm ex}$ and $\tau$
parameters. The upper frame shows $\chi^2$ values over a ($T_{\rm ex}$,
$\tau$) plane, with the $\chi^2$ minimum at the centre of the image (the
white cross). The white and cyan contours correspond to a 1\% and 40\%
increase over the minimum $\chi^2$. The bottom frame shows the spectrum (black
line), the original model (spectrum without noise; red line), and the spectrum
corresponding to the cyan cross in frame a (dashed cyan line).
}
\label{fig:hfs_chi2}
\end{figure}

Figure~\ref{fig:hfs_chi2} illustrates the parameter degeneracy for one of the
spectra at $\sim$3\% noise level. After first fitting the four parameters (in
this case with the Scipy {\tt fmin} routine), the velocity and $FWHM$ values
were kept constant and the $T_{\rm ex}$ and $\tau$ values were varied. The
plot shows the resulting plane of $\chi^2$ values. The best fits correspond to
a narrow valley, where a 1\% change in $\chi^2$ corresponds to a maximum
uncertainty of several degrees in $T_{\rm ex}$ and more than a factor of two
in $\tau$, and these just in this one 2D plane. For the total $\tau$=0.5, all
individual hyperfine components are relatively optically thin, and $\tau$
thus remains poorly constrained.

\begin{figure}
\sidecaption
\includegraphics[width=8.8cm]{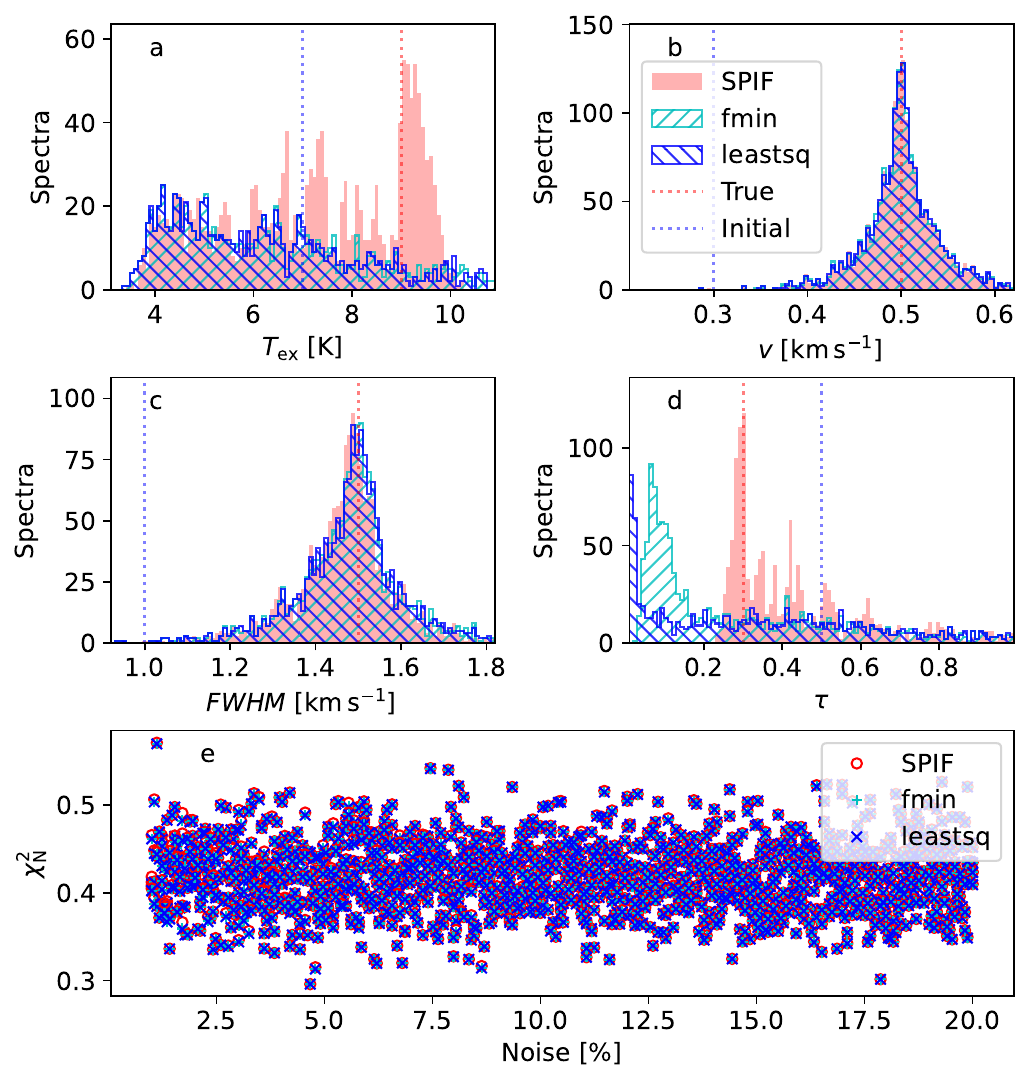}
\caption{
Comparison of parameter estimates from hyperfine fits with SPIF (red
histograms; calculations in single precision) and with Scipy optimisation
(blue histograms). The histograms are based on synthetic observations with a
single velocity component and a relative noise that was varied between 1\% and
20\% of the peak value in the spectrum. The vertical dashed lines indicate the
true values and the initial values values used in the optimisation. The bottom
frame shows the normalised $\chi^2$ values for each of the fitted spectra.
}
\label{fig:hfs_para_cmp}
\end{figure}

Figure~\ref{fig:hfs_para_cmp} shows the distributions of the parameter
estimates and the $\chi^2$ values for all the 322$\times$5 spectra. The SPIF
results are compared to fits done with the Scipy library least-squares
minimisation routine {\tt leastsq} and the general optimisation routine {\tt
fmin} with default tolerances, all fits using the same initial values, as
indicated in the figure. While the $v$ and $FWHM$ distributions are identical,
$T_{\rm ex}$ and $\tau$ show differences. The SPIF calculations were done
using the fastest option (naive gradient descent and single precision). This
results in more sub-structure in the $T_{\rm ex}$ and $\tau$ distributions,
while also the {\tt leastsq} and {\tt fmin} results show some differences. The
normalised $\chi^2$ values (Fig.~\ref{fig:hfs_para_cmp}e) and the
corresponding spectrum profiles (not shown) are nevertheless almost identical.
If the SPIF runs were carried out using the Simplex method or if the fits were
repeated a few times with different initial values, the resulting parameter
distributions of SPIF would be close to that of the {\tt fmin} runs.

\begin{figure}
\sidecaption
\includegraphics[width=8.8cm]{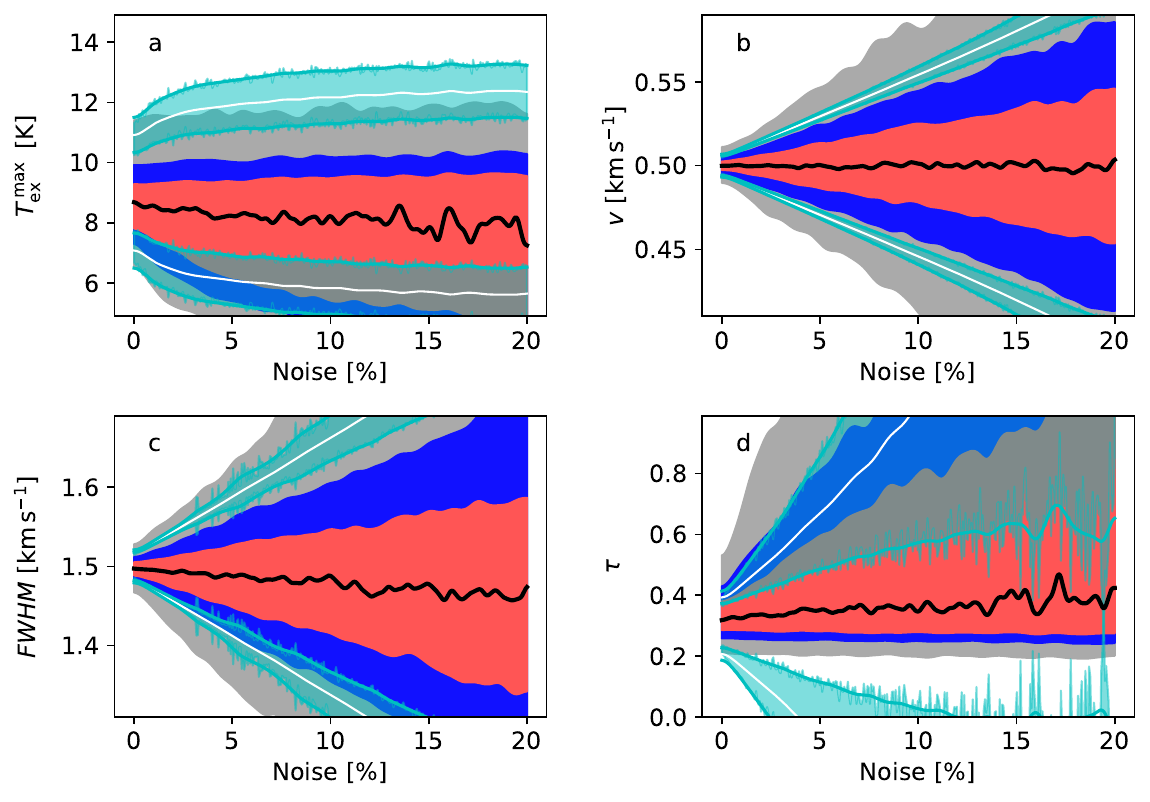}
\caption{
Distribution of best-fit parameters and their error estimates in the case of
synthetic N$_2$H$^{+}$ spectra from the cloud simulation. The sample consists
of 100 spectra for each of the 322 noise levels. The black lines show the
median parameter estimates, and the shaded red, blue, and grey areas
correspond to the [20, 75], [10, 90], and [0.1, 99,9] percentiles of the
parameter estimates, respectively. The median Monte Carlo error estimates are
shown with solid white curves, and the 1-$\sigma$ variation of the error
estimates (between spectra at the same noise level) with cyan shaded regions.
The error estimates are plotted symmetrically relative to the true parameter
values.
}
\label{fig:5a}
\end{figure}

While the results of the three routines are not significantly different, the
different clustering of the results clearly has the potential to bias the
Monte Carlo error estimates. If the $\tau$ values were even smaller, the
optimisation could stop close to the initial values, leading to severe
underestimation of the uncertainty. The use of randomised initial values could
again help to obtain more realistic error estimates, even when individual fits
showed some dependence on the initial values. Figure~\ref{fig:5a} shows the
distributions of the best-fit parameters (322 noise levels with 100 noise
realisation each) and the error estimates from 100 Monte Carlo samples per
spectrum. For each Monte Carlo sample, the fit was repeated ten times with
initial values sampled from normal distribution with 30\% dispersion around
the best-fit values. The error estimates are accurate for $v$ and $FWHM$, and
they are of the correct magnitude also for $T_{\rm ex}$ and $\tau$. The
procedure has the added benefit of also partly addressing the potential
problem of multiple local minima. The Monte Carlo method is still in
relative terms more time-consuming, and in this case (32~200 spectra and $100
\times 10$ separate fits for each), the calculations took some 2.5 minutes of
wall-clock time.

\begin{figure}
\sidecaption
\includegraphics[width=8.8cm]{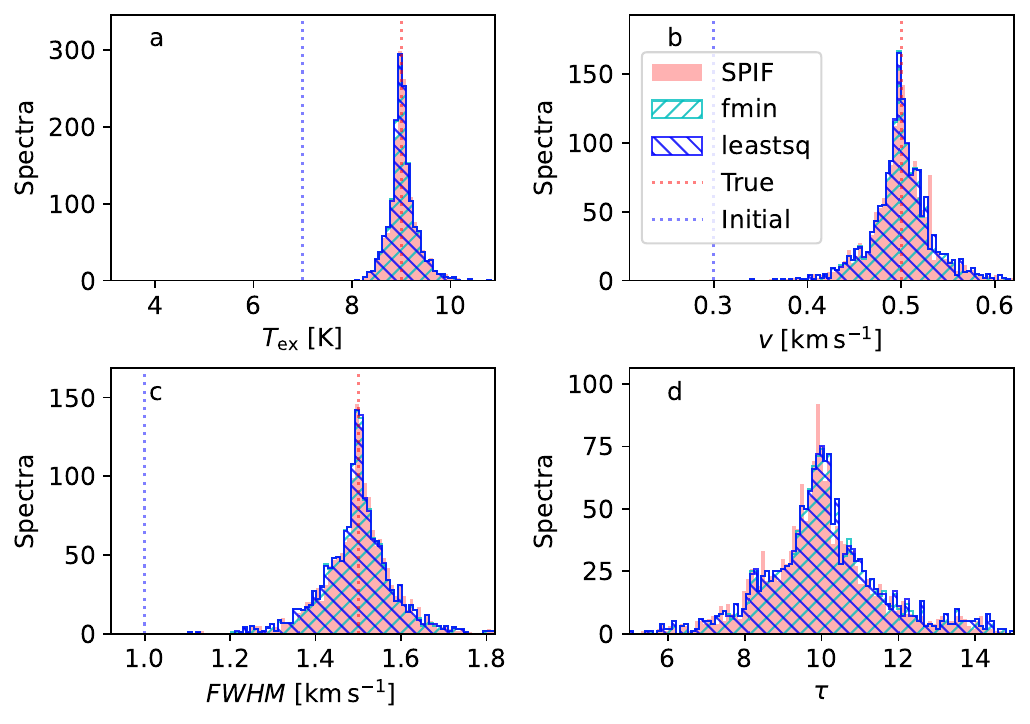}
\caption{
Comparison of parameter estimates from hyperfine fits with SPIF and with Scipy
libraries. The figure is similar to Fig.~\ref{fig:hfs_para_cmp} but for
spectra with the higher total optical depth of $\tau=10$.
}
\label{fig:hfs_para_cmp_hitau10}
\end{figure}

The low optical depth of the above example highlighted the potential problems
that parameter degeneracy might cause in error estimation. Similar to 
Fig.~\ref{fig:2-hfs_colden}, we tested a higher optical depth also in the case
of simple single-component synthetic spectra. The results are shown in
Fig.~\ref{fig:hfs_para_cmp_hitau10}. The fits correspond to the same 1-20\%
noise range as in Fig.~\ref{fig:hfs_para_cmp}, and the only difference is the
higher optical depth, $\tau$=10 compared to $\tau=0.5$. This results in better
than 1\,K accuracy for $T_{\rm ex}$, while the fractional error in the optical
depth can still amount to some tens of per cent at the higher noise levels.

\section{Discussion} \label{sect:discussion}

We have presented a new, GPU-accelerated SPIF program for spectral line
fitting. Its main characteristic is a relatively good computational
performance, which allows the use of repeated fits with random initial values,
in order to increase the chances of finding the globally optimal fit. Error
estimation with Monte Carlo or MCMC methods is also feasible, even for maps
containing thousands of spectra. The performance of the program and some
general properties of spectral fits were investigated using synthetic
observations.

\subsection{Performance of the SPIF program}

We were able to fit 10$^5$-$10^6$ spectra per second with Gaussian models
(Fig.~\ref{fig:timings}). However, these numbers depend on many factors, such
as the number of velocity channels, the hardware used, the selected initial
parameter values, and the number of fitted velocity components. In
Fig.~\ref{fig:timings}), the spectra had only 120 channels, the optimisation
was done with the fastest option in SPIF and a powerful GPU cards. The more
accurate Simplex optimiser was slower by a factor of a few, but still
processing $\sim 10^5$ spectra per second. The use of penalty functions had
only a minor effect on the run times. The third optimiser, the conjugate
gradient method, is equally fast, but can be used only when analytical
derivatives are provided (including those of the penalty functions).

\subsection{Importance of initial values}

The tests highlighted the importance of good initial values.
Figure~\ref{fig:fails} showed examples of how a fit can fail by converging to
a wrong local minimum, even when spectra contain only two perfect Gaussian
components. The situation is usually more complex, as spectra may contain
several non-Gaussian velocity components and extended background components of
indefinite shape (cf. Fig.~\ref{fig:spectra}). The testing of alternative
initial values does not help with the model errors (i.e. the observed emission
not being the sum of, for example, a few perfect Gaussians) but will increase
the probability of finding the global $\chi^2$ minimum for the selected model.
This is important in multi-component fits, where there are more possibilities
for the individual components to converge towards different spectral features.

In the test of Fig.~\ref{fig:chi2}, repeats with random initial values
improved fits noticeably (more than 20\% reduction in $\chi^2$) in about one
third of the cases. The fraction increased slightly with the increasing number
of fitted components, but showed no consistent dependence on the noise level.
The numbers of course depend on the complexity of the observed spectra and
also on the way the initial values are selected.

\subsection{Model errors}

If the observations do not match the assumed model, even a technically optimal
fit will result in an unbiased view of the emission. We investigated how well
the Gaussian fits represent the emission, the line area and velocity
dispersion, in the case of the synthetic observations.

In Fig.~\ref{fig:areas}, the analysis of spectra from the MHD model, there was
a tendency to underestimate the line areas with the largest errors exceeding
50\%. When spectra contain multiple completely separate velocity components,
the fit will of course convert the area of one component. Thus, one needs to
both select the correct number of component and be successful in fitting them.
By choosing the best of the alternative 1-3 component fits (based on AIC) the
results were on average unbiased, and the relative uncertainty of the line
area was similar to the relative noise in the observed spectra. This is of
course not a general rule, since it also depends on how many channels the
spectra span. A higher noise could also cause systematic underestimation, as
weaker emission components are lost in the noise. In the case of
Fig.~\ref{fig:areas}, no such bias was observed. Line area is also a
relatively robust parameter. Even when spectra are skew or contain
sub-structure, the fitted Gaussian can still provide an accurate estimate of
the line area.

In contrast, velocity dispersion $\sigma_v$ is more sensitive to emission at
large velocity offsets, and the fit of a small number of Gaussians is likely
to correspond to a much lower velocity dispersion.  This may also be
desirable, when the study concentrates on a component or a few components, and
the rest of the emission is unwanted line-of-sight confusion. In
Fig.~\ref{fig:dispersion} we examined spectra where a single-component fit
seemed to be appropriate. However, the single-component Gaussian fits were
biased also for this sample, and the velocity dispersion was sometimes
underestimated by more than a factor of two. Such errors are very relevant in
studies of the cloud energy balance, resulting in large uncertainty in the
kinetic energy, $E_{\rm kin} \propto \sigma_v^2$.

\subsection{Error estimation}

Monte Carlo simulation provides a good way to estimate the formal uncertainty
and the correlations between the fitted parameters, even when the model
includes priors (such as those entered via penalty functions). The method may
be feasible even for large samples of millions of spectra, at least for crude
error estimates. The noise of the error estimates themselves increases towards
the tails of the error distribution, which are therefore much harder to
quantify. Figure~\ref{fig:4a} showed results for fits of a single Gaussian,
using only 100 Monte Carlo samples per spectrum. The plotting of [0.01,99.9]
percentile interval was possible only because these corresponded to the
average over 322 noise realisations. The 100 samples per spectrum may be
enough to quantify the error distribution up to $1-\sigma$ level, but the
estimation of the [0.01,99.9] percentile range would require 2-3 orders of
magnitude more Monte Carlo samples. For example, with one million spectra, 10 000 Monte Carlo samples
per spectrum, and 10$^5$ fits per second, the computations would still take
more than one full day. The tails of the error distribution (more
realistically up to 2-3$\sigma$ levels) are still of some interest, because
they may show asymmetries, probability of large deviations, and other
deviations from the normal distribution (e.g. Fig.~\ref{fig:4a} and
Fig.~\ref{fig:4b}).

We showed that MCMC is a potential alternative for error estimation. This was
true at least in the Gaussian fits, even when using the most straightforward
Metropolis algorithm. While Monte Carlo method requires one full optimisation,
a single MCMC step needs only the evaluation of the $\chi^2$ function, and
both methods have similar run times. The only concern is the good mixing of
the MCMC chains, especially for models with a larger number of parameters and
when parameters exhibit very different dynamical ranges. These may require
more complex MCMC algorithms, such as the Robust Adaptive MCMC
\citep{Haario2001} or the Hamiltonian MCMC \citep{Betancourt2011}.  Based on
preliminary tests, these are however not competitive in very simple cases (such
as fits of 1-2 Gaussian components). 

One must also exercise some caution in the use of Monte Carlo and MCMC
estimates. In models consisting of two Gaussians, the fitted components were
rearranged so that they appear in velocity order in each Monte Carlo sample.
In MCMC calculations, the chains that follow different velocity components can
at any point switch identity and continue to follow a different component,
especially when the noise is high. This can render the direct chain-averages
meaningless, although this did not seem to be a problem in
Fig.~\ref{fig:mcmc_2g}. One should therefore register MCMC samples of a
derived quantity that does not depend on the chain identity, such as the total
line area (sum over all components, irrespective their order) or directly the
total optical depth calculated for each MCMC step. This would thus also
directly provide the error distributions for the derived physical quantity.

Compared to the Gaussian fits, the hyperfine fits to synthetic N$_2$H$^{+}$
spectra were more challenging.  All parameters cannot be determined accurately
either on the optically thin or the optically thick limit. We analysed some
optically thin synthetic spectra that showed strong degeneracy between the
$T_{\rm ex}$ and $\tau$ parameters (Fig.~\ref{fig:hfs_stat3},
Fig.~\ref{fig:hfs_chi2}). The degeneracy affects the fits also in a technical
sense, as different fitting routines, tolerances, and initial values may all
lead to slightly different results (Fig.~\ref{fig:hfs_para_cmp}). The run
times were somewhat longer than for the Gaussian models
(Fig.~\ref{fig:timings}), but just in relation to the larger number of 
velocity channels in the N$_2$H$^{+}$ spectra. If optical depths are large,
radiative transfer effects also mean that the observed spectrum does no longer
exactly match the fitted model. This is not restricted to non-Gaussian line
profiles, as spatial variations in excitation conditions, combined with the
different optical depths of the components, can also changes in the hyperfine
ratios \citep[e.g.][]{GolzalezAlfonso1993}. If $T_{\rm ex}$ and $\tau$ cannot
be both determined, one option is to fix one of the parameters to a likely
value, or to use priors to further constrain the solution \citep[cf. Appendix
A in][]{Pyspeckit22}

We did not carry out any multi-component hyperfine fits. These are more
difficult as optimisation problems and may require further priors to avoid
unphysical solutions. As noted in \citet{Juvela2022_ngVLA}, already in a
two-component model there is the risk that some of the $T_{\rm ex}$ values
approaches $T_{\rm bg}$, which would have a vanishingly small contribution to
the modelled intensities but a potentially arbitrarily large contribution to
the estimated column densities. These cases need to be excluded, either by
using suitable priors or as part of the selection of the best model. The
problem does not appear if the components share the same $T_{\rm ex}$ value,
but this is not generally a well justified assumption.

\subsection{Selection of the best model}

In the spectrum analysis, one critical step is the selection of the best model
and especially the correct number of fitted components. This can be decided by
analysing the spectral shape prior to fits \citep[e.g.][]{Gausspy19},
interactively by visual inspection \citep[e.g.][]{Scousepy16}, or by comparing
the statistics of completed alternative fits. 
The last option could be based on the reduced $\chi^2$ values or the
use of AIC \citep[in the present paper and in][]{Clarke2018} or BIC
\citep{Rigby2024}.
\citet{Sokolov2020} used the Bayesian analysis, which incorporates 
user-defined priors into the decision process. There the model selection was
then based on the ratio of the Bayesian evidences, the joint probability due
to likelihood and priors, marginalised over the parameters.

All statistical criteria have their limitations. The observations do not
necessarily follow normal statistics, due to observational imperfections that
cause deviations from normal statistics (i.e. noise plus baseline errors) and
in particular due to model errors, when the spectra contain emission that
cannot be accurately described by the chosen model. The Bayesian approach can
be powerful, but may also sometimes bias the results towards our expectations.
Furthermore, the statistics (such as the basic $\chi^2$) depend on the number
of channels that do not contain signal but are still included in the analysis.
The probability of spurious features also increases as the number of channels
is increased \citep[cf. Appendix A in][]{Gausspy19}. On the other hand, the
exclusion of all channels that do not have significant emission in an
individual spectrum will cause more or less bias, depending on the sensitivity
of the observations \citep{Yan_2021}. Features that are insignificant in an
individual spectrum may be significant at the map level. The information from
the larger scales (e.g. averaged spectra with higher S/N) can lead to more
robust selection of the relevant channels and the number of fitted velocity
components. The inspection of spatially averaged emission is part for example
of the SCOUSEPY analysis \citep{Scousepy19}. In details, the selection of the
optimal model is also likely to depend on the goals of the analysis. The
determination of line area and velocity dispersion have different
requirements. In kinematic analysis, it is preferable to avoid random jumps
that would be caused by the number of fitted components changing from spectrum
to spectrum. Thus, the number of components may need to be fixed over a
certain area and common priors need to be used to ensure consistency of
multi-component fits. Such analysis is not directly built into the SPIF
program but is still partly enabled by it. One possible scheme is to start by
fitting spectra convolved to a lower angular resolution (to increase S/N and
the spatial correlations) and use those results as direct priors for the fit
at the full resolution.

\subsection{Comparison to other tools}  \label{sect:comparison}

A number of tools exist for spectral line fitting, and these may also cover
related tasks, such as noise estimation and the selection of the optimal
number of velocity components to be fitted. Some tools also make it possible
to take into account spatial correlations between spectra.

Pyspeckit has an extensive library of spectral fitting tools including
specific models for studying the hyperfine structure. Here, the user can
define the region of interest either manually or using the built-in GUI. The
initial guesses for decomposing the spectra are input parameters, while the
program utilises the neighbouring spectra to get initial assumptions for a
more efficient fitting. Pyspeckit supports a range of data types and has
inbuilt plotting routines. The versatility of the program helps in easily
integrating it into different pipelines. For example, SCOUSEPY is another
multi-component spectral fitting tool which uses the interactive fitting
routine in Pyspeckit \citep{Scousepy19}. The program allows to targeting of
localised regions compartmentalised into several spatially averaged areas
(SAA). The corresponding spectra from these user-defined SAAs are obtained
using the framework of Pyspeckit. Compared to Pyspeckit, the SPIF program
concentrates more on the fitting step only and does not include, for example,
specific routines for multi-transition fitting of hyperfine spectra.

GaussPy is an autonomous Gaussian decomposition (AGD) algorithm. It requires
one to first transform the data into a format, where each spectrum has its own
independent and dependent spectral arrays. The AGD algorithm uses a smoothed
spectrum and its higher-order derivatives to find the local maxima and minima
to isolate the signal peaks. Users can define one or two (in the case of
two-phase decomposition) parameters, $\alpha_1$ and $\alpha_2$, which are the
regularisation parameters for the smoothing.
We initially fitted a subset of $16^2$ the spectra with GaussPy using an
arbitrary $\alpha$ value. The obtained fitting parameters were used to create
synthetic Gaussian spectra with a fixed noise level, to train the AGD
algorithm to obtain a more accurate $\alpha$. 
As $\alpha$ is a measure of the data smoothness and noise suppression, the
fitting routine becomes slightly more complicated for data sets with a wide
range of noise or component separation (Appendix~\ref{app:stat2}).

\begin{figure}
\sidecaption
\includegraphics[width=8.8cm]{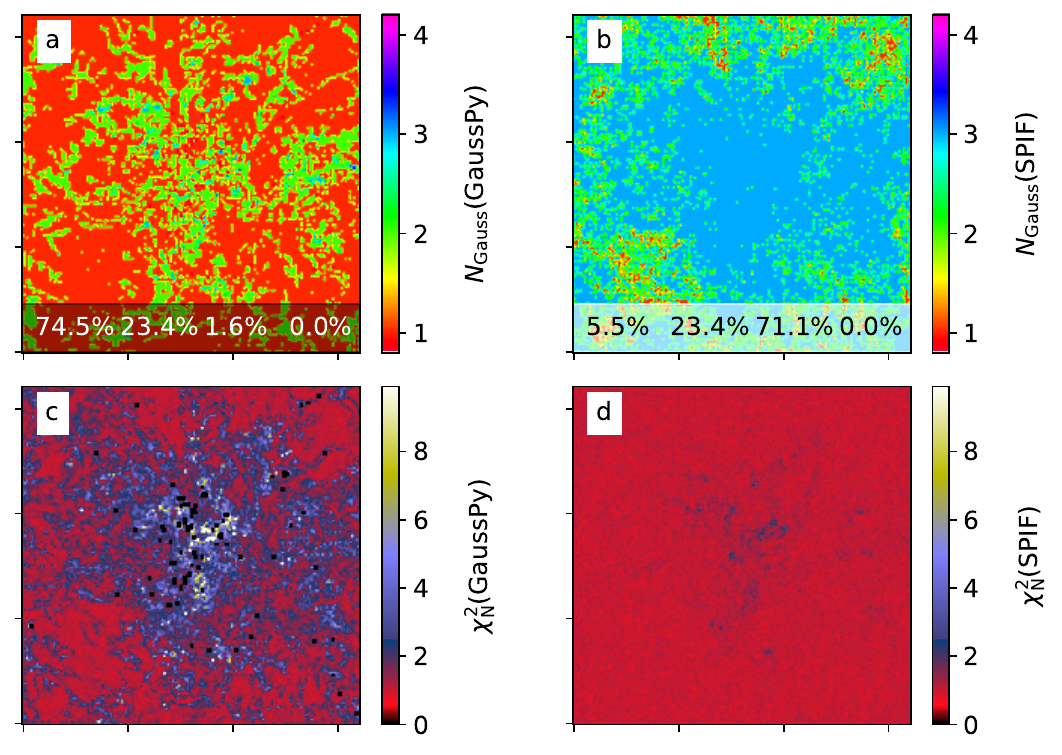}
\caption{
Comparison of GaussPy (left frames) and SPIF (right frames) fits of the
C${18}$0 spectra of the MHD model. The map resolution is $R=16$ and the view
direction $x$. The upper frames show the number of Gaussian components
selected by the programs, with the fraction of spectra fitted with one to four
components listed within the frames. The lower frames show the corresponding
$\chi^2_{N}$ values. 
}
\label{fig:stat}
\end{figure}

Figure~\ref{fig:stat} shows a comparison between GaussPy and SPIF, when
fitting the C$^{18}$O spectra from the MHD model with $\sigma(T)=0.05K$
observational noise. 
The number of GaussPy components (Figure~\ref{fig:stat}a), corresponds to the
smoothing parameter $\alpha$ that was obtained from a trained AGD. 
We fitted the 161$^2$ spectra with GaussPy using nine CPUs in approximately 50
minutes. Compared to the SPIF results with the AIC criterion\footnote{We do
not consider the use of AIC as part of the SPIF program itself, only as one
possible criterion that can be applied based on the alternative fits.},
GaussPy produces many more single-component fits and the resulting fit
residuals are higher (Fig.~\ref{fig:stat}c).

We carried out the GaussPy fits initially with an arbitrary $\alpha$ value
(approximately equal to the channel width, $\delta v$). In this case the fit
ignored many of the minor velocity components seen in multi-component spectra.
With an $\alpha$ obtained from a trained AGD, the ratio of multi-component
spectra has increased. With a larger number of spectra in the training set,
the accuracy of the $\alpha$ obtained from the trained AGD might be further
improved, especially in the case of multi-component spectra.


\begin{figure}
\sidecaption
\includegraphics[width=8.8cm]{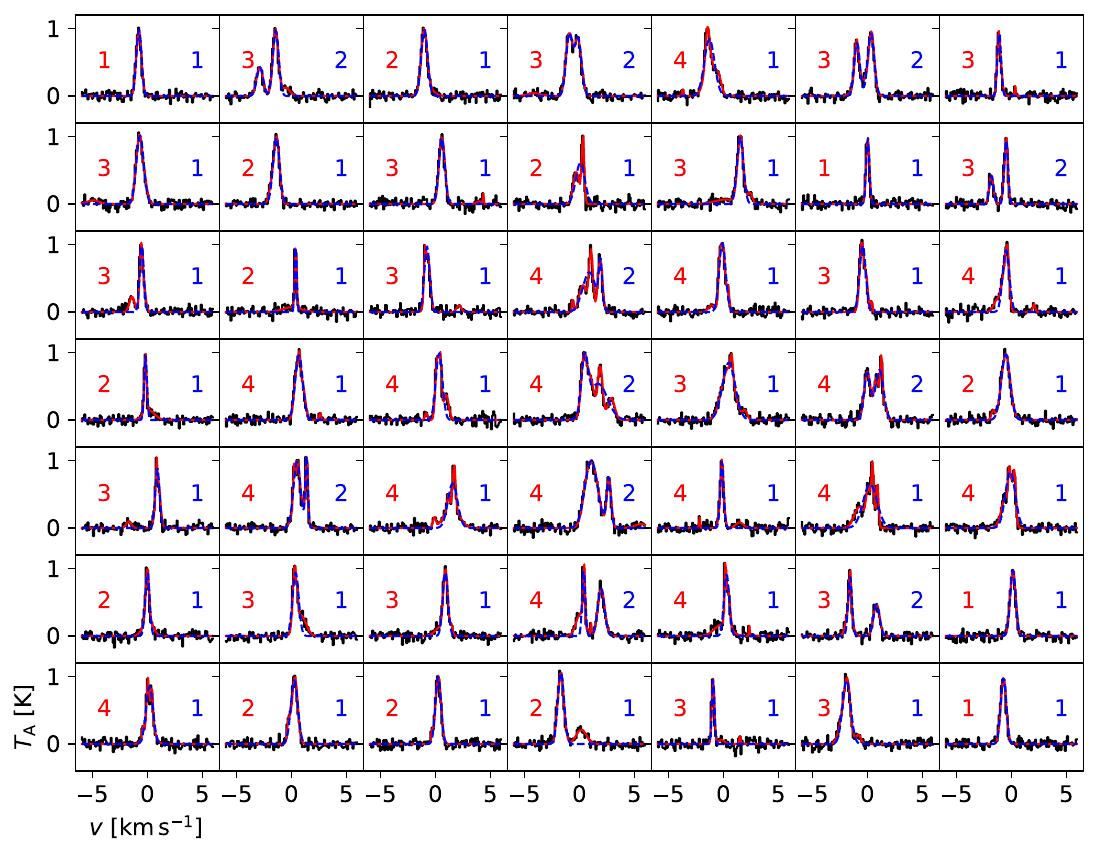}
\caption{
Sample C$^{18}$O spectra from the MHD model, taken at equidistant
positions over the area shown in Fig.~\ref{fig:stat}. The black line shows the
synthetic observations with noise equal to $\sigma(T)=0.05$\,K. The solid red
and dashed blue curves correspond to the SPIF and GaussPy fits, respectively.
The number of Gaussian components selected by the methods are indicated in the
frames with numbers of the corresponding colour.
}
\label{fig:plot_dt_C18O_N5_spe}
\end{figure}

Figure~\ref{fig:plot_dt_C18O_N5_spe} shows as an example 7$\times$7 spectra
that are extracted uniformly over the area shown in Fig.~\ref{fig:stat}. As
indicated by the previous figure, the AIC criterion tends to choose a larger
number of components, which then also leads to lower $\chi^2$ values. In some
cases, one would clearly choose fewer components by eye, although it is
difficult to judge by mere visual inspection the significance of weak features
close to the noise level or small deviations from nearly Gaussian line shapes.
In contrast, GaussPy is more conservative regarding the number of components,
and some spectra appear to remain underfitted. As mentioned earlier, the
``correct'' number of components depends on the use case and the goals of the
analysis, which may thus also necessitate some fine-tuning of the criteria.

\section{Conclusions}   \label{sect:conclusions}

We have presented  the program SPIF for the fitting of spectral lines. The
analysis of its performance and the general tests in fitting synthetic
observations with different spectral models have led to the following
conclusions.

\begin{enumerate}
\item SPIF compares favourably to other fitting routines in computational
speed. The use of modern GPUs allows the fitting of simple models (such as 1-2
Gaussians) at a rate approaching 10$^6$ spectra per second.
\item SPIF run can include retries, where the fit is repeated with different
initial values and the solution of the lowest $\chi^2$ value is chosen. This
was found to improve the results for a large fraction of the cases, although
the typical reduction in $\chi^2$ was only at 10\% level.
\item The selection of the optimal number of fitted velocity components is not
part of the SPIF program. However, as done in the present paper, one can
quickly run alternative fits and use afterwards statistical (and other)
criteria to choose the most appropriate one.
\item The error estimates derived by Monte Carlo and partially with MCMC
method were found to be adequate and also feasible up to samples of tens of
thousands of spectra. However, in some cases (such as optically thin lines
with hyperfine structure) the error estimates might become dependent on the
adopted initial values, and further iterations may be needed to validate the
estimates.
\item The above points contribute to the robustness and the relative 
ease of use of the SPIF program. Thanks to the brute-force approach (the use
of a large enough number of retries with random initial values), the default
input values will usually result in fits that are nearly optimal.
\end{enumerate}

\begin{acknowledgements}
The work was supported by the Research Council of Finland grant 348342.
\end{acknowledgements}

\bibliography{my.bib}

\begin{appendix}

\section{Statistics of the synthetic test spectra} \label{app:stat}

We discuss below some properties of the spectra obtained from MHD cloud
simulation. Figure~\ref{fig:stat} shows the general statistics of the
C$^{18}$O and $^{13}$CO spectra. A large fraction of the spectra are
multimodal. The histograms in Fig.~\ref{fig:stat}a,e show the number of
distinct spectral peaks, where the individual peaks are required to be
separated by a dip of at least 10\% relative depth, in the spectra containing
no observational noise. We exclude individual peaks below 0.1\,mK, to avoid
potential spurious peaks that could be caused by numerical errors. In the full
2573$\times$2573 pixel maps ($R$=1), the highest number of peaks is about ten,
but these correspond to only a fraction of $\sim10^{-5}$ of all spectra. For
larger values of $R$, which also imply convolution with larger beams,
multimodal spectra are in relative terms equally common and are rare only due
to the smaller total number of spectra. Depending on the assumed observational
noise, the synthetic observations should thus often be fitted using multiple
velocity components. 

The large absolute values of skewness (Fig.~\ref{fig:stat}c,g) are also due to
the presence of multiple velocity components. However, even among the spectra
classified as having only a single significant peak, one in four spectra shows
at least moderate skewness ($|\rm skewness|>0.5$). The fraction tends to be
smaller for $^{13}$CO spectra and larger $R$ values, but drops below 10\% only
for stronger spectra above 10\,mK. For normal distribution, the skewness and,
as calculated in Fig.~\ref{fig:stat}, the kurtosis should both be zero. Based
on the central limit theorem, one could have expected that a large beams
(larger $R$) would reduce the kurtosis and the absolute values of skewness.
However, as in the case of the number of peaks, the histograms differ mainly
just due to the smaller number of spectra. Only the largest kurtosis values
are more noticeably reduced as $R$ is increased.

\begin{figure}
\sidecaption
\includegraphics[width=8.8cm]{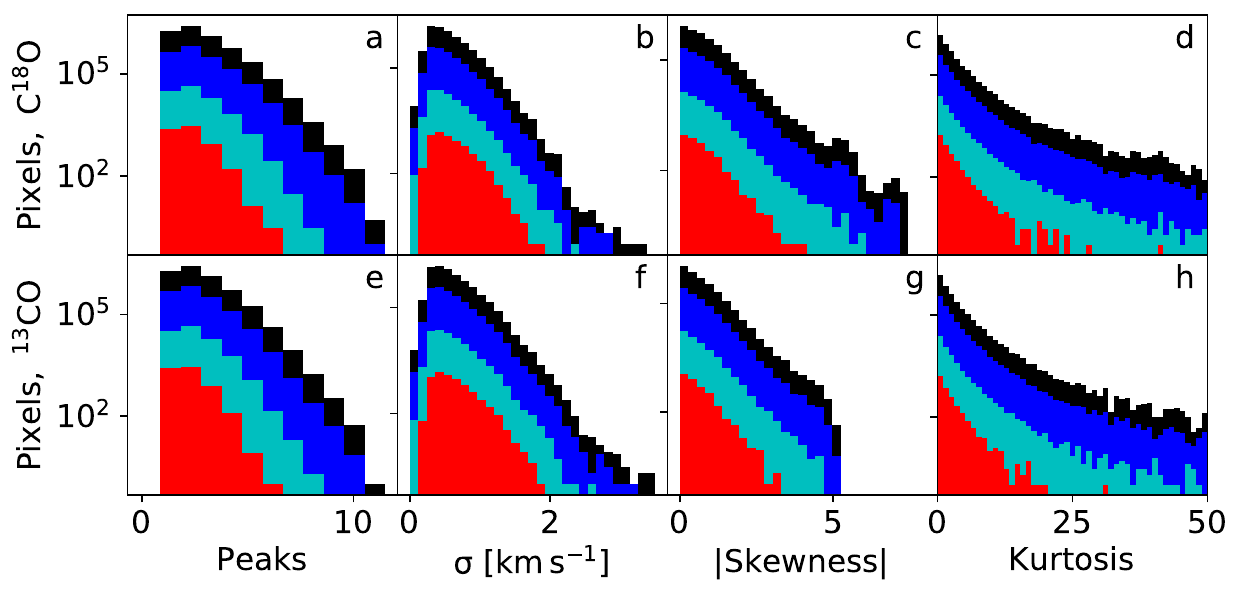}
\caption{
Statistics for synthetic C$^{18}$O (upper row) and $^{13}$CO (bottom row)
spectra. Histograms are plotted for the number of spectral peaks, the line
width converted to standard deviation $\sigma$, the skewness, and the
kurtosis. Each frame contains four histograms that correspond to $R$=1, 2, 8,
and 32, in decreasing order of the histogram height.
}
\label{fig:stat2}
\end{figure}

\section{Notes on the usage of the SPIF program} \label{app:manual}

We provide here a short overview of the input parameters of SPIF, including
the model definition, the use of penalty functions and priors, and the
generation of initial values for the fits. A more detailed description can be
found in the a pdf document at the project GitHub
page\footnote{https://github.com/mjuvela/ISM}. The input parameters related to
the fitted model are provided in a single ini file, and there are also some
command-line parameters that are concerned more with the technical details of
the calculations

The ini file lists names of the FITS files than contain the spectra, with a
separate file for their error estimates. The fitted model is described using
pre-defined {\tt GAUSS} and {\tt HFS} names, for fits of isolated Gaussian
profiles and hyperfine spectra, respectively. The free parameters of the
models are listed explicitly as {\tt x[0], x[1]}, etc. One can fit
simultaneously spectra from up to two files (e.g. different transitions of the
same species), and the use of the same parameter name for both introduces a
explicit constraint between the two fits. In the model definition, the term
{\tt y1} refers to the spectra in the first (and in this case the only) input
FITS file, and the term {\tt v1} refers to its channel velocities. Thus, a
minimal ini-file for fitting a single Gaussians to one set of spectra is
\begin{verbatim}
fits1  =  13CO.fits
dfits1 =  13CO_err.fits                
y1     =  GAUSS(v1, x[0], x[1], x[2])     
prefix =  result
\end{verbatim}
The parameters {\tt x[0]-x[2]} are the peak temperature, central velocity, and
the FWHM value. The fit of spectra from two input files, forced to have the
same radial velocity ({\tt x[1]}), could be
\begin{verbatim}
fits1  =  13CO.fits                    
dfits1 =  13CO_err.fits                
y1     =  GAUSS(v1, x[0], x[1], x[2])  
fits2  =  C18O.fits                    
dfits2 =  C18O_err.fits                
y2     =  GAUSS(v2, x[3], x[1], x[4])
\end{verbatim}

Initial values for the fit of each spectrum can be read from an external file.
However, they can also be set via the ini-file, either to some constant value
or as parameters computed from the spectra. The possible computed parameters
include the peak intensity ({\tt tmax}), the velocity of the peak location
({\tt vmax}), and the FWHM ({\tt fwhm}) computed over the full spectrum. One
also needs to specify, if the calculation is to use the spectra of the first
or the optional second spectrum file ({\tt y1} vs. {\tt y2}). As a example,
when fitting one set of spectra with two Gaussian components, the ini-file
could include lines
\begin{verbatim}
y1    =  GAUSS(v1, x[0], x[1], x[2]) + \
         GAUSS(v1, x[3], x[4], x[5])
init  =  y1:tmax    y1:vmax      y1:fwhm   \
         y1:tmax/2  y1:vmax+0.5  1.2
\end{verbatim}
The model contains two independent Gaussians and six free parameters, and the
initial values are specified in the same order. For the first Gaussian they
are directly the values calculated based on the observed spectra, and the 
second Gaussian uses the same value with some modifications. Only the initial
$FWHM$ value of the second Gaussian is set directly to a constant value of
1.2\,km\,s$^{-1}$.

The ini-file can contain penalty functions and priors. While the penalty
function can be an arbitrary c-language expression involving some free
parameters, the priors are assumed to be properly normalised probability
density functions (e.g. in MCMC calculations). As an example, a penalty for
negative values of the free parameter {\tt x[0]} could be included as
\begin{verbatim}
penalty =   (x[0]<0.0) ? (-x[0]/0.1) : 0.0
\end{verbatim}
For bias terms, SPIF includes a couple of pre-defined functions, and for
example a normal-distributed prior {\tt x[1]}$\sim N(2.0, 0.3)$ can be
entered as
\begin{verbatim}
bias  =   NORMAL(x[1]-2.0, 0.3)
\end{verbatim}
The penalty and bias expressions can also refer to values that are read from
auxiliary FITS files, one value per spectrum. One possible use case is to set
a prior for the fitted velocity based on velocities found in a previous fit of
a different line. Further details on the use of external files and other
ini-file options can be found on the GitHub page.

\section{Synthetic spectra with two Gaussians} \label{app:stat2}

\begin{figure}
\sidecaption
\includegraphics[width=8.8cm]{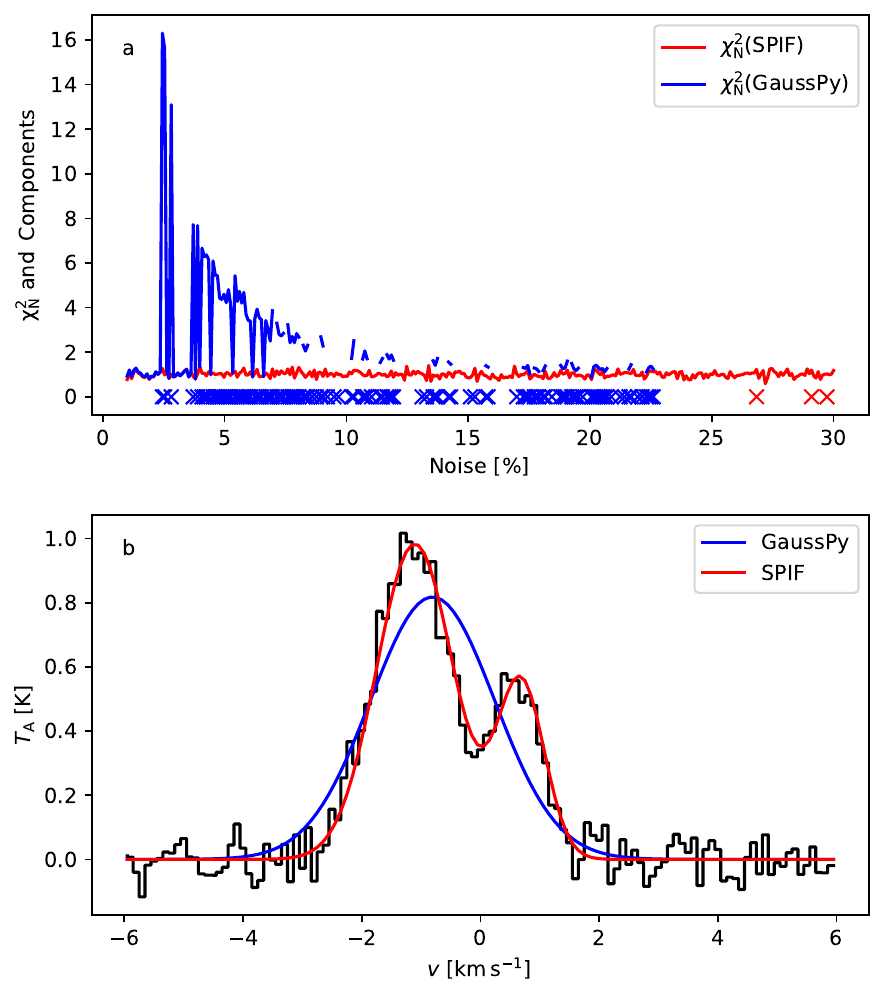}
\caption{
Fit to spectra containing two Gaussian components and with noise varied from
1\% to 30\%. The upper frame shows the reduced $\chi^{2}$ values for GaussPy
(blue line) and SPIF (red line) fits. The crosses indicate cases where
programs have preferred the single-component fit. Some GaussPy fits have
failed, and in those cases no $\chi^2$ value is plotted. In the case of
GaussPy, we modelled the parameter using smoothing parameters $\alpha \sim \delta
v$ for noise $\le 16\%$ and $\alpha \sim 40 \times \delta v$ for noise $>
16\%$. A fixed smoothing parameter for varying noise levels prevents GaussPy
from isolating the multiple components. In the case of SPIF, both one- and
two-component fits were performed and the model with the lower AIC values is
included in the figure. The lower frame shows one example where GaussPy (blue
line) has opted for a single-component fit, resulting in a noticeably higher
$\chi^2$ value in frame a.
}
\label{fig:gausspy}
\end{figure}

Figure.~\ref{fig:gausspy} compares the results of the spectral modelling with
GaussPy and SPIF for a synthetic two-component spectrum of size $322 \times
5$. Here, we have introduced a noise level varying from $1\%$ to $33\%$. We
have used GaussPy single-phase fitting for two chosen values of $\alpha$ for
two parts of this data cube. For the lower noise levels ($<16\%$), we use an
$\alpha$ approximately close to the channel width, whereas for the higher
noise level spectra ($>16\% $), we fix the $\alpha$ to be around $40$ times
the channel width. GaussPy is successful in isolating the two components at
the lower noise levels. As the noise level increases, it becomes more
challenging to fit the multiple components. For chosen values for $\alpha_1$ and
$\alpha_2$ even the two-phase decomposition routine failed to capture the
multiple components at higher noise levels. While a trained smoothing
parameter at different noise levels will better fit the two-component spectra,
the widely varying noise levels can still be non-trivial in fixing the degree
of smoothness.

\end{appendix}

\end{document}